\documentclass[12pt]{emulateapj}
\usepackage[mediumspace,mediumqspace,Gray,squaren]{SIunits} 
\usepackage{natbib}
\usepackage{color}
\usepackage{amsmath}

\newcommand{\ie}{i.\,e.}

\newcommand{\eg}{e.\,g.}

\newcommand{\teff}{$T_{{\rm eff}}$}  
\newcommand{\kms}{\mbox{km s$^{-1}$}}

\newcommand{\vsini} {$v$\,sin\,$i$}

\newcommand{\snr} {\mbox{SNR}}

\newcommand{\vmicro}{\mbox{$\xi_{\rm t}$}}
\newcommand{\gfeh}{\mbox{$[{\rm Fe}/{\rm H}]$}}

\newcommand{\logg}{\mbox{log~{\it g}}}
\newcommand{\loggf}{\mbox{log~{\it gf}}}

\received{22 May 2013}
\revised{18 September 2013}
\accepted{17 October 2013}
\shorttitle{Atmospheric Parameters and Metallicities for M\,4}
\shortauthors{Malavolta et al.}

\begin{document}

\title{Atmospheric Parameters and Metallicities \\ for 2191 stars in the Globular Cluster M\,4}

\author{
Luca Malavolta\altaffilmark{1,4},
Christopher Sneden\altaffilmark{2},
Giampaolo Piotto\altaffilmark{1,4},
Antonino P. Milone\altaffilmark{3},
Luigi R. Bedin\altaffilmark{4},
Valerio Nascimbeni\altaffilmark{1,4}
}

\altaffiltext{1}{Dipartimento di Fisica e Astronomia, 
Universit\`{a} di Padova,
Vicolo dell'Osservatorio 3, 35122, Padova, Italy;
luca.malavolta@studenti.unipd.it;
giampaolo.piotto@unipd.it;
valerio.nascimbeni@unipd.it}

\altaffiltext{2}{Department of Astronomy and McDonald Observatory,
The University of Texas, Austin, TX 78712, USA;
chris@verdi.as.utexas.edu}

\altaffiltext{3}{Research School of Astronomy and Astrophysics, 
The Australian National University,
Cotter Road, Weston, ACT, 2611, Australia;
milone@mso.anu.edu.au}

\altaffiltext{4}{INAF - Osservatorio Astronomico di Padova,
Vicolo dell'Osservatorio 5, 35122, Padova, Italy,
luigi.bedin@oapd.inaf.it}

\begin{abstract}
We report new metallicities for stars of Galactic 
globular cluster M\,4 using the largest number of stars ever observed at 
high spectral resolution in any cluster.
We analyzed 7250 spectra for 2771 cluster stars gathered with the VLT
FLAMES+GIRAFFE spectrograph at VLT.
These medium resolution spectra cover by a small wavelength range,
and often have very low signal-to-noise ratios.
We attacked this dataset by reconsidering the whole method of 
abundance analysis of large stellar samples from beginning to end. 
We developed a new algorithm that automatically determines the 
atmospheric parameters of a star.
Nearly all data preparation steps for spectroscopic analyses are 
processed on the syntheses, not the observed spectra.
For 322 Red Giant Branch stars with $V \leq 14.7$ we obtain a
nearly constant metallicity, $<\textrm{\gfeh}> = -1.07$ ($\sigma$ = 0.02).
No difference in the metallicity at the level of $0.01 \textrm{dex}$
is observed between
the two RGB sequences identified by
\cite{Monelli:2013us}. 
For 1869 Subgiant and Main Sequence Stars $V > 14.7$ we obtain
$<\textrm{\gfeh}> = -1.16$ ($\sigma$ = 0.09) after fixing the
microturbulent velocity. 
These values are consistent with previous studies that have performed
detailed analyses of brighter RGB stars at higher spectroscopic resolution
and wavelength coverage.
It is not clear if the small mean metallicity difference between brighter
and fainter M\,4 members is real or is the result of the low signal-to-noise
characteristics of the fainter stars.
The strength of our approach is shown by recovering a metallicity close to 
a single value for more than two thousand stars, using a dataset that is
non-optimal for atmospheric analyses.
This technique is particularly suitable for noisy data taken 
in difficult observing conditions.
\end{abstract} 


\section{Introduction}\label{sec:intro}

Galactic globular clusters contain $10^4 - 10^6$ stars, yet until the 
1970's they generally were considered to be simple systems with members
born at one epoch from chemically homogeneous gas.
Their actual complexity has been revealed as photometric and 
spectroscopic data have steadily increased in quality and quantity.
We now know that in most (all?) globular clusters (GCs) there are large 
star-to-star variations in all light elements that are susceptible 
to proton-capture fusion reactions (Li, C, N, O, Na, Mg, Al).
Several clusters have internal variations in the heavy neutron-capture
elements (\eg, Y, Zr, Ba, La, Eu).
The large number of papers that have contributed to these
conclusions have been discussed in several major reviews, \eg,
\cite{Kraft:1979a}, \cite{Freeman:1981a}, \cite{Kraft:1994cf},
\cite{Gratton:2004dy}, \cite{Piotto:2010iu}, \cite{Gratton:2012ej}. 

Recently attention has been drawn to the discovery of multiple
color-magnitude sequences in several GCs, for example  $\omega$
Centauri \citep{Anderson:1997vg,Bedin:2004hn}, NGC~2808 
\citep{Piotto:2007br}, NGC~1851 \citep{Milone:2009fn}, 
47~Tuc \citep{Anderson:2009cr,Milone:2012gs}, NGC~6397
\citep{Milone:2012dr}, 
and many others where we find evidence
of multiple populations in the SGBs \citep{Piotto:2012iz}.
In followup spectroscopic studies these split evolutionary sequences 
have been interpreted in to be due to metallicity variations 
(\eg, M22, \citealt{Marino:2011a}) and helium abundance variations 
(\eg, $\omega$~Cen, \citealt{Piotto:2005ii};
NGC~2808, \citealt{Bragaglia:2010a}).
An increasingly popular interpretation of the metallicity and abundance 
ratio inhomogeneities is in terms of multiple stellar generations
with individual clusters (\eg, \citealt{Carretta:2010il}, 
\citealt{Gratton:2012ej}).
To determine whether the derived abundance variations are
  statistically significant or not, it is essential to attack these questions with large spectroscopic samples of high quality.

Several years ago our group targeted several clusters in spectroscopic 
surveys with the Very Large Telescope (VLT), in order to study their
internal velocity dispersions ($71.D-0205(A)$ and $72.D-0742(A)$ ESO 
Proposal, P.I. Piotto). 
The spectra for these surveys were taken specifically to measure radial 
velocities, and at first glance they may seem not suitable for chemical 
composition analysis: the 
wavelength range is short ($\leq$ 214~\AA), the spectra often are noisy
 (\snr$\geq 10$) in this spectral region (centered at 5250~\AA) many lines
  are intrinsically blended or suffer  for the limited resolving power
  of the instrument ($R\simeq 24000$)

However, the spectra still contain much more information than just 
velocities. 
Given the large number of stars observed (several thousand per cluster) 
and their great luminosity range (from the Upper Red Giant Branch down 
to the Main Sequence) we have tried a fresh approach to see what 
other spectroscopic information might be extracted in these clusters.
We decided to concentrate only on our M\,4 dataset to develop the new
analytical methods.

M\,4 (NGC~6121) is a bright, well-populated, mildly metal-poor
Galactic GC.
It is usually regarded as the GC nearest to the Sun: $d$~= 1.8~kpc 
\citep{Dixon:1993a}, 1.7~kpc \citep{Peterson:1995iy}
and 2.2~kpc \citep{Harris:1996a}\footnote{
http://physwww.physics.mcmaster.ca/~harris/mwgc.dat}.
The closeness of M\,4 makes its upper red-giant branch (RGB) accessible
to high-resolution spectroscopy with medium-large telescopes:
at the red-giant tip, $V$~$\approx$ 11.8, and at the horizontal
branch (HB) $V$~$\approx$ 13.5.
However, M\,4 lies between the Sun and Galactic center and
consequently
suffers significant interstellar extinction, $A_V$~$\sim$ 1.3~mag.
Additionally, this extinction is variable over the cluster face
(\citealt{Cudworth:1990a}, \citealt{Ivans:1999hf},
\citealt{Marino:2008du}, \citealt{Hendrick:2012cw}),
and so transformations of colors and magnitudes into luminosities
and temperatures are not straightforward.
Evidences of multiple generations have been found in the RGB
\citep{Marino:2008du} and HB \citep{Marino:2011a}, and the presence of
two distinct sequences in the RGB has been confirmed by \cite{Monelli:2013us}.

Comprehensive high-resolution spectroscopic investigations of atmospheric 
parameters, metallicities, and abundance ratios in M\,4 red giants have 
been conducted for several decades by several author groups, including
\cite{Geisler:1984a}, \cite{Gratton:1986a}, \cite{Brown:1992a}, 
\cite{Drake:1994a}, \cite{Ivans:1999hf}, and \cite{Marino:2008du}.
The metallicity of M\,4 now appears to be well determined at 
[Fe/H]~$\approx$ $-$1.15 with a range from individual studies 
of $\approx\pm$ 0.15 (\citealt{Kraft:2003a}, \citealt{Gratton:2004dy}, and references
therein).
This cluster has been studied extensively in an attempt to understand 
its blue CN bandstrength bi-modality (\citealt{Norris:1981a}, 
\citealt{Smith:2005a}
and references therein).
There also have been several high-resolution spectroscopic
studies that have focused on star-to-star the behavior of
more limited element groups: Na, Al, and O, \cite{Drake:1992a}; 
Li, \cite{Dorazi:2010a}, \cite{Monaco:2012li}; Na in HB stars, \cite{Marino:2011a}; K, \cite{Takeda:2009a};
neutron-capture elements, \cite{Yong:2008a,Yong:2008b};
He, \cite{Villanova:2012a}. Recent efforts have been
  concentrated in systematic abundances 
analysis of large sample of stars with comparison with the theoretical
yelds of elements to identify the origins of the multiple populations,
\cite{Villanova:2012ls}, \cite{Carretta:2013al}, \cite{DOrazi:2013ag}.
Here is a partial list of questions we hoped to answer with 
our M\,4 spectroscopic data.

\begin{itemize}

\item Can temperatures and gravities derived from spectra be mapped 
reliably onto the color-magnitude quantities? 
Are the scatters in these relationships readily understood in terms
of the uncertainties in both sets of quantities?

\item Is the derived metallicity the same at the RGB tip and the MS?  
Is the star-to-star metallicity scatter compatible with the uncertainties,
or is there evidence for a genuine variation such as seen in M\,22?

\item Can efficient, physically defensible, and automated
algorithms be developed for these data that have general applicability
to other cluster data sets?

\item Is it possible to get interesting abundance ratios from these spectra?

\item Can the velocity information (stellar binarity, overall velocity
dispersion, cluster rotation) derived by \cite{Sommariva:2009cz} be
improved?

\end{itemize}

In this work, we will focus on the parameter atmosphere determination
of M\,4 stars. 
A companion paper (Malavolta et al., in preparation) will discuss derivation
of new, very precise radial velocities in this cluster.

\section{Observations and Data Reduction}\label{sec:obsspec}

The spectra analyzed in this study were originally presented by 
\cite{Sommariva:2009cz} for an investigation of the internal velocity 
dispersion and binary fraction of M\,4.
The data were gathered with the VLT Fibre Large Array Multi Element 
Spectrograph (FLAMES; \cite{Pasquini:2000tk}) employed with the GIRAFFE 
medium-high 
resolution spectrograph in MEDUSA multi-fiber mode.
In this configuration one can obtain single-order spectra for 
132 objects (target stars and sky) in each integration.

The GIRAFFE HR9 setup was chosen, which produces spectroscopic
dispersion of 0.05~\AA/pixel and measured 4-pixel resolving 
power R~$\equiv$~$\lambda/\Delta\lambda \simeq 25.800$, in the 
wavelength range 5143~\AA\ $<\lambda <$ 5356~\AA. 
This spectral region is characterized by a large number of lines, many of 
them blended, and the presence of the strong \ion{Mg}{1}~b triplet. 

The M\,4 target stars were selected from an astrometric and
photometric catalog based on Wide Field Imager (WFI) data from the
ESO/MPIA 2.2m telescope \citep{Sommariva:2009cz}. 
The original selection criterion was that each star had no neighbors 
with $V_{neighbor} < V_{target} + 2.5$  within an angular distance of 
$1.2\, \textrm{arcsec}$ (the fiber radius is $0.6\, \textrm{arcsec}$). 
Since the original radial velocity survey goal was to search for 
spectroscopic binaries, almost all the stars have multi-epoch observations.

For our analysis, extensive photometric data for M\,4 have been kindly
provided by Peter Stetson (private communication) from his
comprehensive catalog of homogeneous broadband photometry for Milky Way 
clusters and dwarf spheroidals.
The basic description of that project is reported in
\cite{Stetson:2000if} and \cite{Stetson:2005bg}.
Photometry for over 70,000 M\,4 stars in broadband $U$, $B$, $V$, $R$, and $I$
was obtained in a multi-year effort involving many telescope/instrument 
combinations, as summarized in \cite{DAntona:2009kg}.
Nearly all of the stars have both $B$ and $V$ measurements, 
about 76\% also have $R$ measurements, but only about 43\% were observed
in the $I$ passband.
Photometry of a subset of nearly 800 M\,4 stars has been made publicly 
available as part of Stetson's ``Photometric Standard Fields''\footnote{
http://www1.cadc-ccda.hia-iha.nrc-cnrc.gc.ca/community/ \\ STETSON/standards/},
and most of those stars are included in our large catalog.
The magnitudes given in the public database are consistent with those
that we are using.

\begin{figure}
\includegraphics[width=\linewidth]{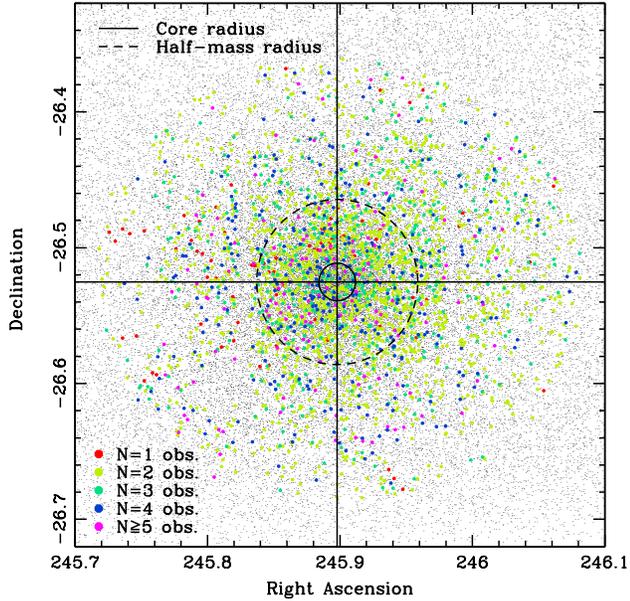}
\caption{The spatial distribution of our targets. The whole sample is
  inside the tidal radius of the cluster (not shown in the plot). Targeted stars are color
  coded according to the number of available spectra. Black points
  are stars from the Stetson (private communication) M4 database that
  were not observed in our survey.}
\label{fig:cmd_map}
\end{figure}

\begin{figure}
\includegraphics[width=\linewidth]{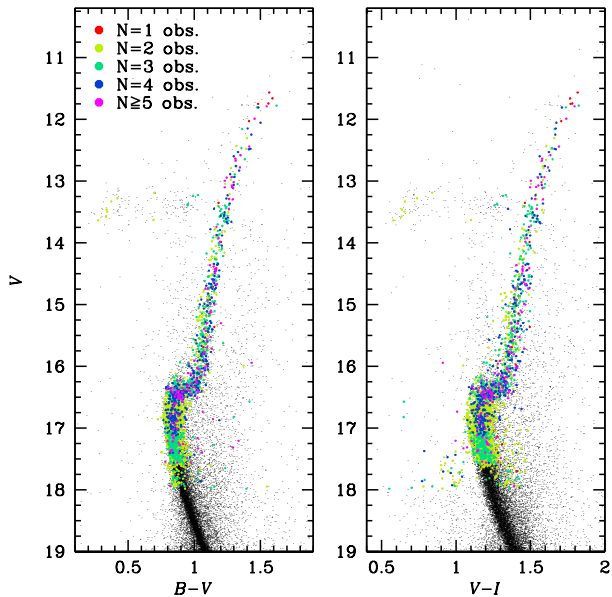}
\caption{Photometric characteristics of our target stars, superimposed
  to the Color-Magnitude Diagram of the cluster. Stars are color-coded
  using the same criteria of Figure~\ref{fig:cmd_map}.}
\label{fig:cmd_color}
\end{figure}

Spatial  properties of our 
target stars are illustrated in Figure~\ref{fig:cmd_map}, where the
targets are color-coded according to the number of spectra available
for a given star.
The cluster core radius (where the surface luminosity has decreased
by a factor of two from the center) and the half-light radius are
also drawn in this figure, indicating that essentially all
regions
of M\,4 has been sampled spectroscopically. 
All the observed stars are inside the tidal radius of the cluster (not
shown in the figures).
Photometric characteristics of the stars are shown in
Figure~\ref{fig:cmd_color}, 
here again color-coded depending on the number of observations as in
Fig.~\ref{fig:cmd_map}. 
The two panels show the color-magnitude diagram in $V/B-V$ (left-hand
panel) and $V/V-I$ (right-hand panel) before any correction for reddening. 
The limiting magnitude of the targets, $V~\lesssim$~17.5, has been set by 
the requirement that a single M\,4 integration had $\textrm{\snr}>10$ for 
each star in a single observation. 
Although Horizontal Branch stars are present in our dataset, they have 
been excluded from the analysis. We think that their higher
temperature respect to the rest of the sample and their evolved status
requires a more careful analysis that is beyond the goal of this paper.

A total of 2771 stars covering color-magnitude diagram positions 
from the upper red-giant branch to about one magnitude fainter than the 
main-sequence (MS) turnoff (TO) luminosity have been observed. 
The spectra span a temporal period of six years, and we refer to 
\cite{Sommariva:2009cz} for a description of the observation strategy. 
We added 306 new spectra obtained in 2009 that targeted MS stars 
already observed in the previous epochs.
Since these data originally were gathered to determine the M\,4
velocity dispersion and to assess the binary-star fraction,
nearly all stars have been observed at least twice, and three or more
spectra have been obtained for nearly 40\% of the sample.
We summarize the observation numbers in Table~\ref{tab-nobs}; a
total of 7250 individual spectra have been used in our study.

\cite{Sommariva:2009cz} used the ESO GIRAFFE standard reduction pipeline 
\citep{Blecha:2000ud} to reduce the M\,4 data from raw CCD exposures to final
extracted, flat-fielded, wavelength-calibrated spectra.
In order to maximize the information content of these spectra, we 
decided to revisit the reductions, paying particular attention
to spectral extraction and wavelength calibration. 
We adapted the \textit{empirical Point Spread  Function} concept from photometry
\citep{Anderson:2000bt,Anderson:2006a} to spectroscopy to obtain a
more accurate determination of the cross-dispersion spread function of the fibres. 
The use of a more accurate profile results in a reduction of the noise
for spectra at low \snr\ when using the optimal spectral extraction
algorithm, as shown by \cite{Horne:1986bg}. 
Wavelength calibration has been performed with an alternative 
technique as well. Precise RV determination from GIRAFFE data is the subject of
a forthcoming paper, and it will not be discussed here. \\

\begin{figure}
\includegraphics[width=\linewidth]{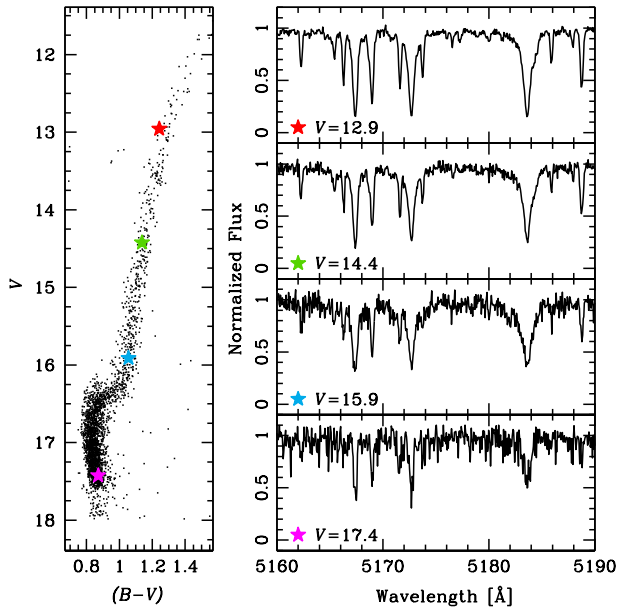}
\caption{Degradation of the \snr\ with magnitude is shown by displaying four
  example spectra of stars with different magnitudes. These spectra
  have been chosen among the best exposed during observing runs with
  good atmospheric conditions and without Moon illumination.}
\label{fig:CMD_NORED}
\end{figure}

Degradation of the  \snr\  with magnitude is demonstrted in 
Figure~\ref{fig:CMD_NORED} by showing representative spectra 
of stars with different magnitudes. 
These spectra have been chosen, among target stars of similar magnitude,
to be the best exposed during observing runs without Moon illumination. 
Other spectra at the same magnitude can have lower \snr\ depending 
on exposure times and observational conditions.

\section{Overview of the Spectrum Analyses}

M\,4 GIRAFFE spectra have been gathered in such a way that each
individual exposure has a minimum \snr greater than 10, since this
value is considered the lower limit to obtain radial velocity with a precision of a 
few hundreds m/s with this instrument \citep{Sommariva:2009cz}. 
However, atmospheric stellar parameter and abundance derivation at 
stellar fluxes that are comparable to the noise level requires very careful 
handling of the spectra.
The usual procedure is to process the observed stellar flux as much as 
possible in order to be comparable with the synthesis: sky contribution 
is subtracted, spectra are corrected by flat-fielding, normalized and
finally rebinned in a linear wavelength scale.
Every step in this procedure will produce an increase or an unintended 
correlation in the noise level of the observed spectrum.
Fortunately, this problem is usually negligible when working on spectra 
with good \snr, but it can lead to a significant loss in precision in 
stellar atmospheric parameter determination when the \snr\ is very low 
This is exactly the situation we face with the majority of our observed 
spectra.

After many numerical experiments with our data, we adopted a new
approach to handle the spectrum preparations.  
Specifically, after extraction and wavelength calibration of the raw 
spectra, we decided not to perform any subsequent data corrections.  
We made this choice in order to avoid introduction of any false 
correlations into the spectrum noise. 
All the steps that usually lead to sky-corrected, rest-frame, 
continuum-normalized stellar spectra were instead performed in the 
opposite direction:  deliberate 
degradation
of the ``perfect'' 
synthetic spectra to best match the real extracted spectra.  
Each of these steps is described in detail in the next few sections.
However, application to each spectrum was done in a near-simultaneous
fashion, so here we give an outline of our overall procedure, referring
to the sections in which each procedure is developed.

In the following, we use the term \textit{exposure} to refer to a single    
observation for a given star. 
For every exposure of the star under study, the synthesis to be compared 
is subject to these transformations:  
\begin{itemize}
  \item the synthesis is moved in the radial velocity space to match        
    the measured total radial velocity shift of the exposure (\ie, star     
    heliocentric motion plus the Earth motion);  
  \item the level of sky flux, composed of a scattered solar (moonlight)
    spectrum and a \textit{gray} continuum are determined 
    (Section~\ref{sec:sky_flux});
  \item the appropriate continuum slope function for the exposure
    is derived (Section~\ref{sec:cont_norm_flux}):
  \item preliminary stellar atmospheric parameters are deduced from
    photometric data (Section~\ref{sec:atmos}):
  \item synthetic spectra appropriate to these parameters are computed
    (Section~\ref{sec:spectral-synthesis}).
  \item both synthetic and solar spectra are degraded to the  
    resolution of the instrument, determined from the Th-Ar exposures, 
    using a Gaussian kernel;
  \item the synthetic spectrum, solar spectrum and gray continuum 
    are rebinned in the wavelength scale of the exposure, keeping constant 
    the amount of flux for unitary wavelength;
\item the rebinned synthesis and sky spectra are multiplied for their       
    continuum function, and together with the rebinned gray continuum         
    they are linearly combined using as coefficient the corresponding 
    flux constants, as determined in Section~\ref{sec:sky_flux}  and
    Section~\ref{sec:cont_norm_flux}.
\item the resulting spectrum is multiplied for the flat-field
  of the fiber used to gather the data to include instrumental effects, such as the
 characteristic efficiency curve along an order (\textit{blaze} function) in echelle
  spectrograph.
\end{itemize}  
We call the result of these operations the \textit{processed synthesis} to      
distinguish it from the raw synthesis described in 
Section~\ref{sec:spectral-synthesis}.
Since every exposure differs from the others due to different observational 
conditions, variations in the instrument response, Moon illumination etc. 
this process must be performed for every exposure of a star, and any 
exposure can be compared only with the relative processed synthesis.       
As a final comment here, stars in GCs are known to be slow rotators, so   
the effects of possible rapid rotation are not considered in computing
the the processed synthetic spectra.
Nevertheless it would be very easy to incorporate this effect in the 
analysis, as well as determining the best fitting  \vsini\ value 
during the $\chi^2$ minimization.

\section{Sky flux determination}\label{sec:sky_flux}

For every GIRAFFE exposure on the M\,4 targets, about 5-10 fibers
have been allocated to empty sky regions. Sky subtraction is usually
performed by coadding the spectra from these fibers and subtracting
the result from the observed spectrum of the star. There are several drawbacks when
doing so: 
\begin{itemize}
\item These spectra are intrinsically very noisy, having the sky a
  maximum brightness of $\simeq 17.5\, \textrm{mag} / \textrm{arcsec}^2$
  at $30 \,
  \degree$ from full Moon. Even when coadded, the resulting spectrum is
  still very noisy, then increasing significantly the noise of the 
  sky-subtracted spectra especially in those cases where star and sky
  fluxes are comparable. 
\item Each fiber has a different transmission efficiency. 
  The error in its determination is propagated when the sky spectra have 
  to be coadded, and when the coadded spectrum is subtracted from the star
  spectra.
\item Sky spectra must be rebinned to a common wavelength scale to be
  coadded, then either the science spectra must be rebinned to this common
  scale or the coadded sky spectrum must be rebinned in the
  science spectra wavelength scale, in order to perform the sky
  subtraction correctly. The rebinning of a low \snr\ spectrum can introduce correlated noise and
  should be avoided. 
\item several fibers must be allocated for the sky spectra, reducing
  the efficiency of a multiplexing instrument such as GIRAFFE.
\item even if the allocated sky fibers are scattered around the field
  of view, the coadded spectrum wil not be sensitive to any spatial
  variability of the sky and the individual spectra are too noisy and
  their number too low to determine it over the  $\simeq 500\,
  \textrm{arcmin}^2$ covered by a MEDUSA plate.  
\end{itemize}

\begin{figure}
\includegraphics[width=\linewidth]{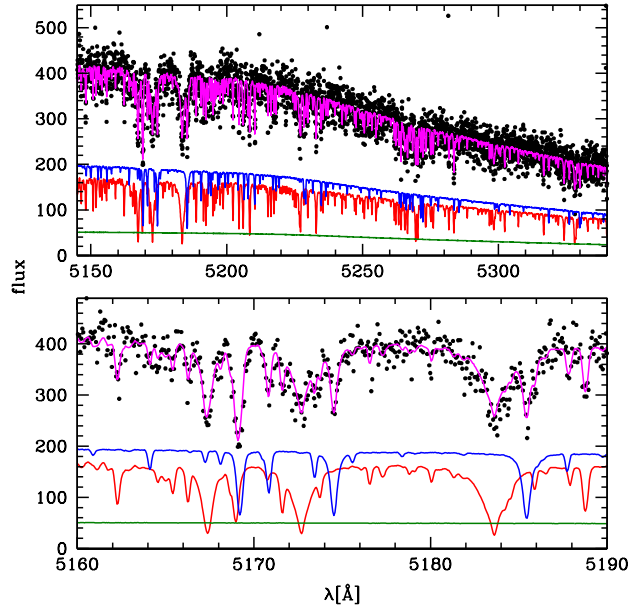}
\caption{Observed spectrum of a star with $V=17.73$, obtained a few days
  before full Moon. 
  The synthetic sky spectrum (red line), stellar spectrum
  (blue line) and a grey continuum (green line)
here shown with instrumental effects already applied, 
 are combined (magenta  line) and compared to the observed stellar
 spectrum (black points).  
  The upper panel shows the entire wavelength range of our spectra,
  while lower panel shows the \ion{Mg}{1}~b triplet region, to highlight the 
  contribution of the sky spectrum to the observed total star+sky spectrum.}
\label{fig:spectra_faint}
\end{figure}

Since many of the observing runs occurred in gray/bright Moon time,
the sky spectrum constitutes a significant part of the flux.
An example is given in Figure~\ref{fig:spectra_faint} where the 
observed spectrum of a relatively faint program star ($V=17.73$) is 
decomposed into the stellar (blue line) and the sky (red line) contributions.
The spectrum in the figure was taken on 2006 September 04,
 $\simeq 5$ days before full Moon and at a distance of $\simeq 40
 \,\textrm{deg}$. 
In analyzing the coadded sky spectrum we discovered that the sky light was
composed of two distinct parts: one resembling the solar spectrum, due
manly to the presence of the Moon, and another featureless \textit{gray}
one (green line). 

This gray feature was first noticed in sky spectra taken during dark 
time, \ie\  without Moon illumination, and confirmed to be
necessary when trying to fit the coadded sky spectra from Moon-illuminated 
nights with the solar spectrum. 
Intrinsic sky brightness due to zodiacal light and human light
pollution \citep[see][]{Patat:2008ct} or the presence of scattered light
inside the dome are both plausible explanations. 
We limit ourselves to taking account of this feature, without 
investigating its origin.

One can see clear evidence in Figure~\ref{fig:spectra_faint} of 
sky contamination in the star's low \snr\ spectrum, especially in the 
strong \ion{Mg}{1}~b lines (lower panel). 
We note that fortunately in our spectral range no telluric absorption
lines are present, so no correction for that potential problem is needed. 

The inclusion of the sky in the correct proportion with respect to the stellar
flux is a fundamental part of a good synthesis match of the data. 
To determine the sky flux level, we proceed in the following way. 
The extracted and wavelength-calibrated sky spectra are corrected for
fiber transmission and flat-field normalization. 
Then the amount of sky flux per unit wavelength is determined by integrating 
the sky spectra over 2~\AA\ intervals; this value has been chosen as a good
compromise between reducing the noise and preserving the main
spectral features of the spectra. 

In order to account for the loss of flux due to solar spectral lines, we 
employed the U.S. National Solar Observatory (NSO) flux atlas 
(\citealt{Kurucz:1984ab}, \citealt{Wallace:2011jg})\footnote{
Available at http://diglib.nso.edu/Wallace\_2011\_solar\_flux\_atlas.html}, 
smoothing it to the spectral resolution of our data. 
This solar model is integrated in the same way as described in the preceding paragraph.

The values determined from the observed sky spectrum are fitted with a linear
combination of two functions: a low-order polynomial multiplied by
the values from the solar model, to take into account the Moon
illumination,  and a constant for the gray continuum.  
The resulting functions represent the continuum flux per
unit wavelength as a function of the wavelength of the moon illumination 
and the gray continuum level, respectively. 
The former function is split in two parts: a constant representing the 
average flux of the sky, and a normalized function for the shape.  
We will refer to them as \textit{flux constant} and 
\textit{continuum function} respectively. 
Since the gray continuum is a constant, its continuum
function is simply a unitary constant value 
(before including instrumental effects)

The coefficients of the functions as derived above represent only the
first step in the sky flux derivation. 
During the continuum determination process, described in the next section, 
the flux constants are refined using the sky lines in the spectrum as a 
proxy, to take into account the sky variability and fiber transmission 
differences. 

Precautions to ensure that the extra absorptions in a faint stellar 
spectrum are truly sky contributions and not accidental inclusion of some 
faint non-cluster member star in the fiber apertures have been taken.  
Not only must the mean sky radial velocities are consistent with a rest 
velocity system, but their fluxes positively must correlate with lunar phase. 
We have also rejected individual sky spectra that had anomalously high 
flux, indicative of accidental pointing at a faint star. 
These procedures convinced us of the real nature of the sky contaminations
to our stellar spectra.

\section{Continuum Normalization}\label{sec:cont_norm_flux}

Continuum rectification proved to be challenging due to the
restricted total wavelength range (5140$-$5360~\AA) of our spectra and 
to the presence of the strong \ion{Mg}{1}~b absorption lines in the first 
50~\AA\ of the spectra.
The GIRAFFE HR9 setting was deliberately chosen for the original
\cite{Sommariva:2009cz} study to take advantage of the line richness in
this spectral region.
In addition to the \ion{Mg}{1}~b lines there are many strong \ion{Fe}{1}
lines, which makes this wavelength domain particularly attractive 
for radial velocity studies. 
But this wealth of absorption lines creates many wavelength intervals
of overlapping spectral features, so defining an adequate number of 
points for a \textit{global} continuum normalization is a very difficult task.
Local normalization, usually determined by visual inspection when
measuring equivalent widths, is also unsuitable for two reasons: the
large number of spectra in our sample; and the subjectivity in manually
selecting the continuum (which probably would have introduced a 
systematic bias in our lower \snr\ spectra). 
Spectral synthesis requires good normalization in the whole wavelength 
range of interest, and while is still feasible to manually adjust the 
continuum to match the synthesis when the number of stars is low, for
our very large sample we had to find an alternative way.

A first estimate of the continuum stellar flux per unit wavelength is 
determined in a similar way as done for the sky flux. 
The sky and grey continuum contributions are removed from the 
observed spectra. 
Then for each star we adopt a synthetic spectrum 
with photometric atmosphere parameters (Section~\ref{sec:atmos}) and
provisional metallicity $\textrm{\gfeh}=-1.15$ (Section~\ref{sec:intro}) to 
calculate the \textit{flux constant} and \textit{continuum function}
relative to the stellar component of the spectrum, following the same
steps described in Section~\ref{sec:sky_flux} for the sky
continuum.

The synthetic stellar spectra and the solar spectra are degraded to match the
spectral resolution of the instrument  and together with the unitary
function for the gray continuum they are rebinned into the
wavelength scale of the science spectrum, keeping constant the flux
per wavelength unit. 
All spectra are then multiplied for the corresponding 
continuum function and the resulting spectra are linearly combined, using the 
flux constants as coefficients. 
The instrumental effects (\eg\ multiplication for flat-field) are
finally applied to make the  comparison with the observed spectrum possible. 

The flux constants are determined by fitting the resulting
model spectrum to the observed one by $\chi^2$ minimization 
(Equation~\ref{eq:spec_chi2}), in the same way as we do to determine the
atmosphere stellar parameters (Section~\ref{sec:abun_an}).
Small adjustments in the continuum functions also are allowed, to deal 
with spectral regions affected by the wings of strong spectral lines, 
\eg\ in the proximity of the \ion{Mg}{1}~b triplet. 
 
A binary mask is included in the $\chi^2$  determination to exclude
points too sensitive to variations of temperature, gravity,
microturbulence and metallicity. 
This mask is derived by taking two boundary values for each stellar 
parameter (\ie\  the photometric value $\pm$ its estimated error; see
Section~\ref{sec:atmos}) and flagging those points that show a percentage 
variation over a fixed threshold. 
A detailed explanation is given in Section~\ref{sec:abun_an}. 
Points that have normalized flux lower than $0.8$ in the stellar 
synthesis are excluded too, while the lines in the sky spectrum are 
retained to allow a better determination of the sky flux level.

The flux continuum functions and the flux constants are adjusted as described
above for each set of stellar atmosphere parameters used during the
$\chi^2$ determination, so that the derived values will not be
influenced by the initial choice in the photometric parameters for
the continuum determination. 
A good starting point however allows a quick recalibration of this function. 

Radial velocity shifts due to the intrinsic velocity of the star and the Earth motion 
are applied to the stellar synthetic spectrum. 
No rebinning of the observed spectra is necessary, and no
noise is introduced since all the operations are performed on noise-free 
high-resolution synthetic spectra.

\section{Stellar Atmospheric Parameters from Photometry}\label{sec:atmos}

In every minimization algorithm, a good starting point always results
in a faster determination of the variable of interest. 
Spectral synthesis is a very lengthy process by itself, and with four 
parameters to be determined by $\chi^2$ minimization and several thousands
of stars to be analyzed, this task can be excessively time-consuming. 
Greater speed in this process can be provided through initialization
of the process with photometrically 
derived stellar parameters, but some precautions are necessary. 

M\,4 is known to have both a large mean reddening, $E(B-V)=0.34$
\citep{Harris:1996a}, and large differential reddening across its
surface.  
We have adopded the latest determination of overall M\,4 parameters derived
 by \cite{Hendrick:2012cw}; these are listed in Table~\ref{tab-parameters}.

We have corrected
the $B/(B-V)$ CMD for differential
reddening using the technique described in \cite{Milone:2012dr};
Here we summarize their method. 
The reddening direction in the CMD is defined as 
$\theta={\rm atan} {(A_V/E(B-V))}$. 
A subsample of the photometric catalog is identified by manually excluding 
those stars that lie clearly outside the cluster sequences, 
and including only those stars in the RGB, SGB and TO, since the angle
between the reddening direction and the sequences is maximum (the
reddening direction is almost parallel to the MS).
A fiducial line along the RGB, SGB and TO is manually determined and then for 
every star in the catalog the median shift in the reddening direction with
respect to the fiducial line of the 35 closest stars of the subsample
is computed. This value is taken as indicative of the amount of the
differential reddening, taking advantage of the fact that stars in a
local spatial region (a few $\arcmin$) of M\,4 all should be shifted 
by the same amount.

\begin{figure}
\includegraphics[width=\linewidth]{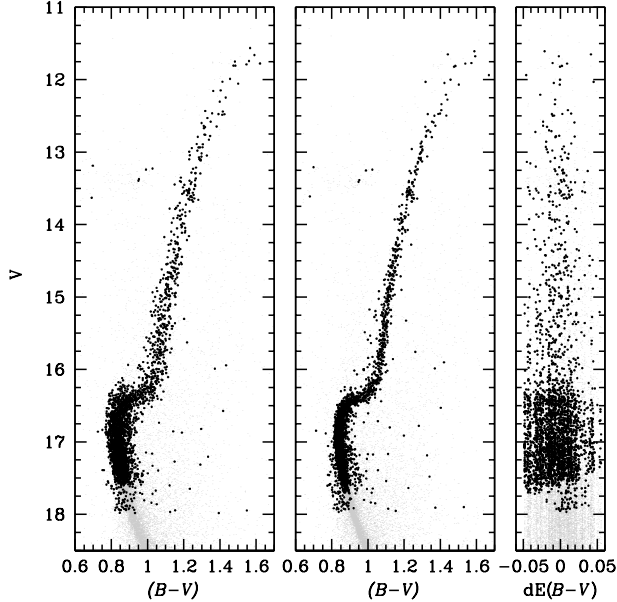}
\caption{The $V/(B-V)$ diagram before (first panel) and after
  (second panel) the differential reddening correction, with the
  cluster stars in gray and target stars highlighted in black. The
  third panel shows the amount of the correction.  
  The overall mean reddening correction has not been applied yet.}
\label{fig:cmd_bv_red}
\end{figure}

\begin{figure}
\includegraphics[width=\linewidth]{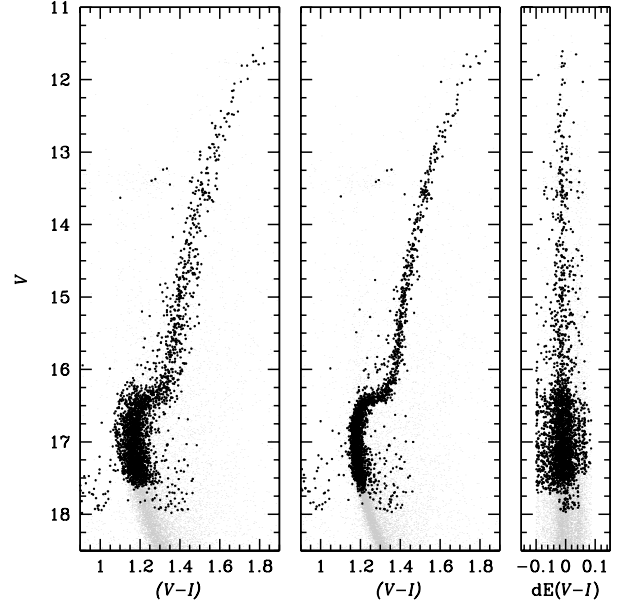}
\caption{As in Figure~\ref{fig:cmd_bv_red}, for the $V/(V-I)$ plane.}
\label{fig:cmd_vi_red}
\end{figure}

Then the $R_V$ value from Table~\ref{tab-parameters} is used to convert 
dE($B-V$) to d$A_{V}$.
In the same way the $V/(V-I)$ diagram has been corrected for
differential reddening, to obtain a second color index whose correction 
is independent from the first one. 
Although one photometric band is in common, the two reddening
corrections can still be considered independent since we are
correcting an effect in color. 
The differential reddening-corrected CMDs in the two color combinations
for our target stars are shown in Figure~\ref{fig:cmd_bv_red} and
Figure~\ref{fig:cmd_vi_red}, along with the amount of correction as a
function of magnitude. 
The overall cluster reddening correction
(Table~\ref{tab-parameters}) has not been applied here yet.

The two color indices have been used to obtain two estimates of the 
(photometric) stellar atmosphere parameters, using the calibrations 
described in the next subsections. 
Although very precise, we must keep in mind
that these value will be used only as input values for our
spectroscopic determination.

\subsection{Effective Temperatures}\label{subsec:photo_teff}

For every star two independent values for effective temperatures are obtained
from $(B-V)$ and $(V-I)$ colors using the relations from
\cite{Ramirez:2005bz} for giant stars and \cite{Casagrande:2010a} for
subgiants and dwarfs stars. 
The switch between the two relations has been set at $V^{switch}=16.1$,
as described below.

\begin{figure}
\includegraphics[width=\linewidth]{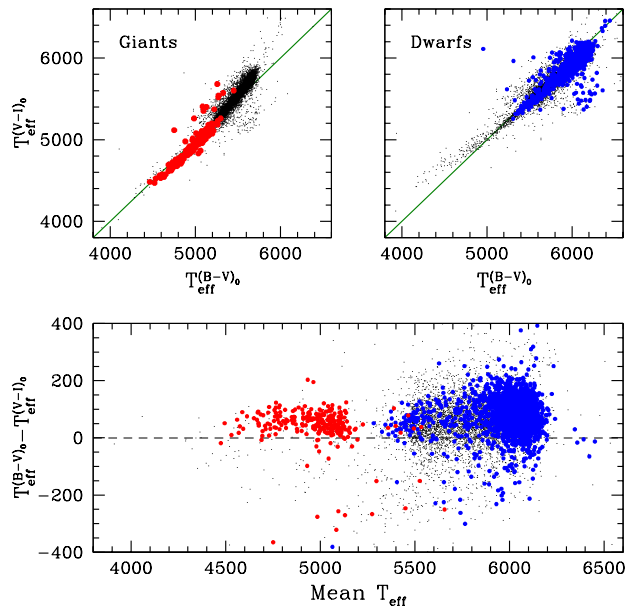}
\caption{Differences in derived effective temperature using the colors
  $(B-V)_0$ and $(V-I)_0$. The upper panels show the temperature
  differences when using the calibration for
  giant (left-hand panel) and dwarf (right-hand panel)
  stars. 
  Equality of temperatures is represented by the green lines.
  Switching the calibration at $V=16.1$ provides the lower
  scatter in the difference when considering giant (red points) and
  dwarf (blue points) stars together (lower panel).}
\label{fig:cmd_atmo}
\end{figure}

The differences in derived effective temperature using the two colors 
are shown in Figure~\ref{fig:cmd_atmo}. 
The two upper panels show direct comparisons between the derived temperatures 
using the two colors, for giants stars in the left-hand panel and 
dwarfs/subgiants in the right-hand one. 
The threshold of the switch has been chosen by looking where the two
temperatures begin to differs systematically, \ie\  the points start to
deviate from the 
green
lines.  
The lower panel shows the difference of the color-derived temperatures
with respect to their mean, which is the value that will be used to derive
the other stellar parameters. 
A gap around $\textrm{\teff} \simeq 5200 \,\kelvin$ is clearly 
visible: \cite{Casagrande:2010a} report a systematic difference
of $\simeq +100 \,\kelvin$ with the calibration of \cite{Ramirez:2005bz} 
for dwarf stars, and probably such difference is affecting also the
calibrations for giant stars. 
We did not attempt to correct this difference, letting the parameter 
determination algorithm solve it.

\begin{figure}
\includegraphics[width=\linewidth]{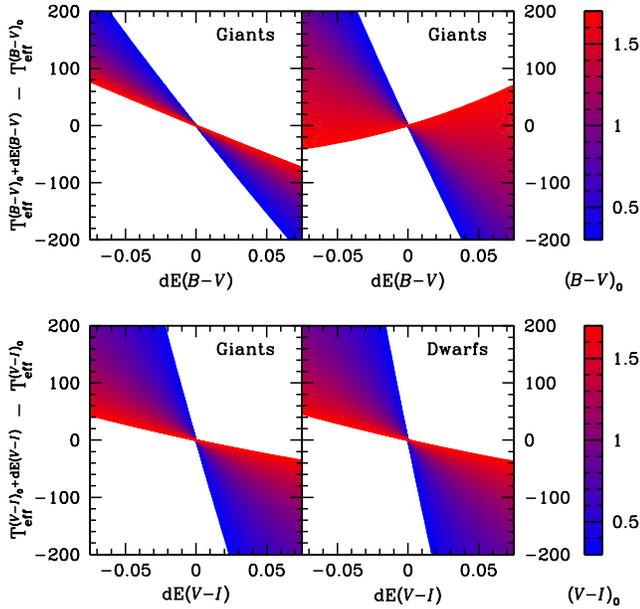}
\caption{The variations in temperature as a function of  dE($B-V$)
   and dE($V-I$) can be used as the corrections in temperature due to 
   the differential reddening correction, or as the error in temperature 
   for a given error in color. 
Values have been determined using a metallicity of $\textrm{\gfeh}=-1.15$.}
\label{fig:teff_dred}
\end{figure}

It is interesting to see the dependence of temperature on small
variations of colors, as given by the calibration. 
Figure~\ref{fig:teff_dred} shows the difference 
between the temperature derived from the cited calibrations when using
a given input color (color-coded according to the bar on the right)
and the temperature obtained after including some color excess
(horizontal axis of the plots) to the considered color. Calculations
have been performed for both giant and dwarf stars, using respectively
\cite{Ramirez:2005bz}  and \cite{Casagrande:2010a} calibrations, for
the two colors at our disposal and a metallicity of
$\textrm{\gfeh}=-1.15$. 
The variations in temperature as a function of  dE($B-V$) and dE($V-I$) 
can be used as the correction in temperature due to the differential 
reddening correction, or as the error associated to temperature for a 
given error in color. 
Note that for a given star the temperature derived using the $(V-I)$
color is in general more sensitive to small variations than using $(B-V)$, but usually the
reddening correction is more reliable when using passbands close to
the infrared. These two effects counterbalance to give similar errors
in the temperatures derived from the two colors.
From the scatter in the difference between the  temperatures
  derived using the two colors , we estimate an average error of  
$\textrm{\teff}=150 \,\kelvin$ for photometric derived temperatures.

\subsection{Surface Gravities}\label{subsec:phot_logg}

Photometric gravities are calculated from the classical relation:   
\begin{equation}\label{eq:phot_gravity}
\begin{split}
 \textrm{\logg}^{\star}  & = \textrm{\logg}^{\odot}+4 \log
 \frac{\textrm{\teff}^{\star}}{\textrm{\teff}^{\odot}} \\ 
 & + 0.4(M_{V, *}+ \textrm{BC} - M_{\rm bol, \odot})+\log
 \frac{M_*}{M_{\odot}} 
\end{split}
\end{equation}
where we have adopted a distance of $1.8 \,\kilo pc$ 
(Table~\ref{tab-parameters}) to relate the reddening-corrected relative 
magnitudes to absolute $V$ magnitudes $M_{V, *}$.
A mean mass of  $M_{\star}=0.85 M_{\odot}$ has been assumed. 
For every star the bolometric correction is obtained interpolating the
table from \cite{Houdashelt:2000hp} for the values of temperature 
and gravity of the star and iterating over $\textrm{\logg}_*$ until 
convergence is reached.
Canonical solar values $\textrm{\teff}^{\odot} =5770K$, $\textrm{\logg}^{\odot}=4.44$,  
$M_{\rm bol, \odot}=4.75$  (see for example \citealt{AllenAstrophysicalQuantities}) have been adopted.

\subsection{Microturbulent Velocities}\label{subsec:phot_microvel}
Initial values of microturbulent velocities for stars cooler than
$5000 \,\kelvin$ are derived using the relation from \cite{Marino:2008du}:
\begin{equation}\label{eq:micro_marino}
  \textrm{\vmicro} = -0.254 * \textrm{\logg} + 1.930
\end{equation}
This relationship has been derived from giants stars in M\,4, but it might
not be suitable for dwarf stars. 
For dwarf stars we initially used the relation from \cite{Gratton:1996va}: 
\begin{equation}\label{eq:micro_gratton}
 \textrm{\vmicro} =1.190*10^{-3} * \textrm{\teff}  -0.900* \textrm{\logg} -2.00
\end{equation}
The two calibrations have similar values at $\simeq 5100 K$, or
$\textrm{\logg}=3.1 ~\textrm{dex}$ 
 
\begin{figure}
\centering
\includegraphics[width=\linewidth]{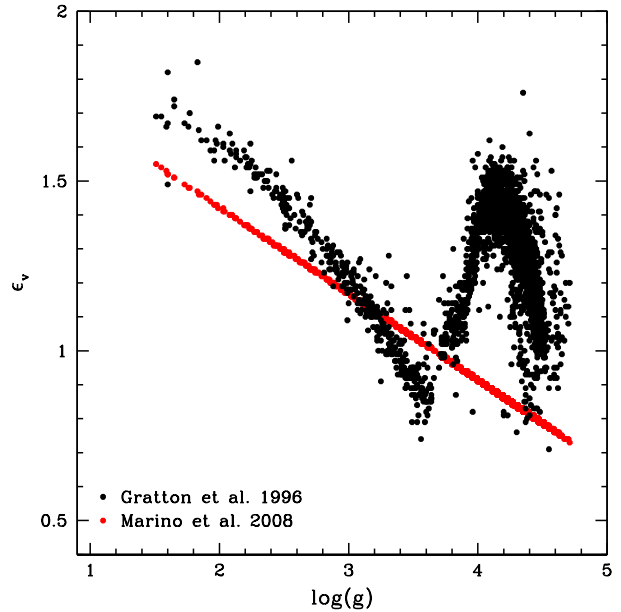}
\caption{Calibration of microturbulence as a function of gravity from
  \cite{Gratton:1996va} (black points) and \cite{Marino:2008du} (red
  points) for our target stars. The first calibration is also
  dependent on temperature, hence the spread in the plot.}
\label{fig:logg}
\end{figure}

As discussed later, several problems have been encountered while
determining the microturbulent velocities of dwarf stars. 
We finally decided to use the velocities computed by 
Equation~\ref{eq:micro_marino}, given the fact that these
values are used only for the first iteration of stellar atmosphere
parameters determination.

\section{Spectral synthesis}\label{sec:spectral-synthesis}

The first ingredient for a  good spectral synthesis is a list of
spectral lines as accurate and complete as possible. 
All the atomic lines in the spectral range 5140$-$5360~\AA\ 
have been collected from several sources in literature, 
see \cite{Kurucz:1995wn}, \cite{Sobeck:2007jb},  \cite{Lawler:2007ke},
\cite{Lawler:2009fn}, \cite{Biemont:2011ie}, \cite{Roederer:2011ac}
and references therein.
The listed values for the  oscillator strength \loggf\ of each line are 
usually the results of laboratory experiments or semi-empirical estimates.  
Calculated \loggf's from theory may be subject to large uncertainties, 
which fortunately can be corrected in numeric experiments with 
astrophysical spectra.

We have used the NSO Solar Atlas (\citealt{Kurucz:1984ab}, 
\citealt{Wallace:2011jg}) and the Atlas of Arcturus Spectrum 
(\citealt{Hinkle:2000wp}) to improve the transition probabilities in our 
line list. 
For the Sun we have assumed the standard atmospheric parameters
$\textrm{\teff} =5777 \,\kelvin$ , $\textrm{\logg} =4.44$, $\textrm{\vmicro} =1.5 \,\kilo\meter
/\second$ , $\textrm{\gfeh}=0.00$. 
For Arcturus there is no general consensus on the atmospheric parameters. 
We have decided to use the values listed by \cite{Massarotti:2007dk}
which are $\textrm{\teff} =4325 \,\kelvin$, $\textrm{\logg}=2.1$, $\textrm{\vmicro}=2.0 \,\kilo\meter
/\second$ , $\textrm{\gfeh}=-0.60$. 
The model atmospheres are generated by interpolation in the 
\cite{Kurucz:1992ab} model atmosphere grid\footnote{
Available at http://kurucz.harvard.edu/grids.html} 
calculated with the ``New Opacity Distribution Function''
(\textrm{ODFNEW}, \citealt{Castelli:2004ti}). 
The syntheses are generated using the current version of the Local 
Thermodynamic Equilibrium (LTE) code MOOG \citep{Sneden:1973el}. 
Oscillator strength values have been modified iteratively until synthetic
spectra produced satisfactory matches with the observed solar and 
Arcturus spectra. 
When it was not possible to find a value to match both spectra satisfactorily, 
priority has been given to the Solar one, given the uncertainty in the 
stellar parameters and chemical composition of Arcturus. 
In some cases it has been necessary to add some atomic lines with arbitrary 
parameters in order to reproduce the solar spectra in places where there
is no known atomic or molecular identification.  
These \textit{fake} lines have been
chosen to be \ion{Fe}{1} transitions with excitation energies of 3.5~eV and
\textit{gf}-values adjusted to match the observed spectral
features. 
While these lines have been excluded in the atmospheric parameter
determination, their inclusion in the synthesis ensure a more reliable
determination of the continuum.

\begin{figure}
\centering
\includegraphics[width=\linewidth]{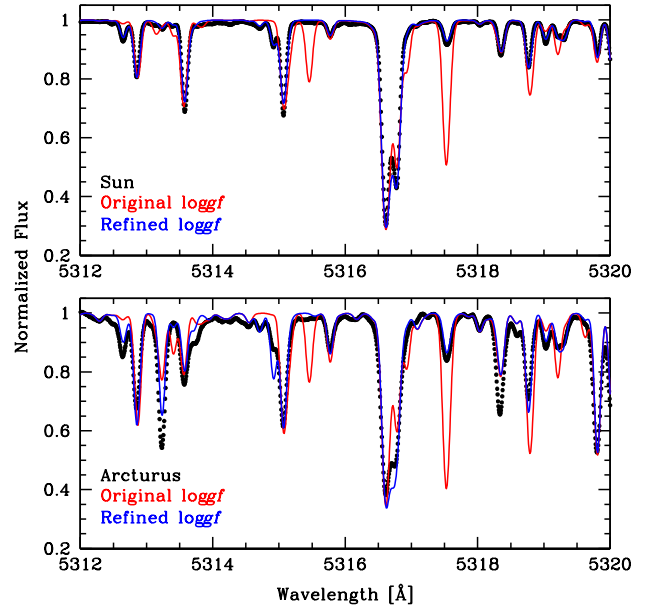}
\caption{The improvement on spectral synthesis from before (red line) and
  after (blu line) the refinement of the \loggf\ values in the input
  line list, compared with high \snr\ spectrum of the Sun (upper panel)
  and Arcturus (lower panel).}
\label{fig:moog_solarc}
\end{figure}

Figure~\ref{fig:moog_solarc} shows the improvement in
the synthesis from before (red line) and after (blue lines) the refinement
of \loggf\ values. 
For the Sun the improved line list causes a reduction of the dispersion of 
the synthesis respect to observed points from $\sigma =0.0507$ to 
$\sigma=0.0147$, while for Arcturus we have a reduction from 
$\sigma =0.0688$ to $\sigma=0.0336$.
Further improvement in the dispersion can be achieved by noting that
Arcturus has a non-solar abundance distribution 
(\eg, \citealt{Peterson:1993eu}, \citealt{Ramirez:2011jk}).
In Figure~\ref{fig:moog_solarc} it is clear that the observed 
spectrum depths exceed the computed ones for transitions near 
5313.2~\AA\ and 5318.3~\AA.
The species responsible for these absorption features are
\ion{Ti}{1} at 5313.24~\AA\ and \ion{Sc}{2} at 5318.36~\AA.
Both Ti and Sc are overabundant by 0.2$-$0.3 dex with respect to
Fe in Arcturus (\citealt{Ramirez:2011jk}), while the synthesis have
been generated using solar-scaled abundances. Accounting for those
overabundances would essentially erase the observed/synthesized 
mismatches at these wavelengths.

The syntheses for M\,4 stars have been generated again using the 
Castelli-Kurucz \citep{Kurucz:1992ab} model atmospheres based on opacity 
distribution functions, this time with $\alpha$-element abundance 
enhancements (\textrm{AODFNEW}, \citealt{Castelli:2004ti}). 
The interpolator code was kindly provided to us by Andy McWilliam
and Inese Ivans.
We modified this code so that the synthesis program automatically set the 
abundance of the $\alpha$ elements to the ones measured by  
\cite{Ivans:1999hf}, namely $\log \epsilon(\mathrm{O})=  7.68 $, $\log
\epsilon(\mathrm{Mg}) =  6.52 $, $\log \epsilon(\mathrm{Si}) = 6.60 $,
$\log \epsilon(\mathrm{Ca}) = 5.12 $, $\log \epsilon(\mathrm{Ti})=
3.79$ for $\textrm{\gfeh}=-1.15$, with those values scaled accordingly
with \gfeh.

Since many stars in our sample share similar temperatures,
gravities and microturbulent velocities, to speed up the
determination of the atmosphere parameters we have created a grid of
synthetic spectra with a spacing of $100 \,\kelvin$ in temperature, $0.1$ in
gravity, $0.1 \,\kilo\meter/\second$ in microturbulent velocity and $0.05$ in
metallicity. The grid is built on the photometric atmospheric parameters
derived for all the stars in the sample. For a specific set of
parameters, the synthesis is computed by linear interpolation of the
syntheses  with the closest parameters within the spacings of the
grid.

\section{Stellar parameters and abundance analysis}\label{sec:abun_an}

In this section we discuss the algorithms used to iteratively determine
\teff, \logg, \vmicro, and \gfeh\ for each M\,4 target star.

Suppose that we have a star with $n_{obs}$ exposures, each one
with  $n_{pixel}$ flux points $F(n,i)$ for the exposure $n$ at the
pixel $i$ and its associated wavelength value. 
For each set of atmosphere parameters we can calculate
the $\chi^2$ as given in Equation~\ref{eq:spec_chi2}:
\begin{equation}\label{eq:spec_chi2}
\chi^2 = \frac{ \sum_{n=1}^{n_{obs}} \sum_{i=1}^{n_{pixels}}  \frac{(F(n,i) -
  S_o(n,i))^2}{S_o(n,i)}   p(n,i) w(n,i) }
{ \sum_{n=1}^{n_{obs}} \sum_{i=1}^{n_{pixels}}  p(n,i) w(n,i) }
\end{equation}
where $p(n,i)$ is a binary mask to include only points sensitive to
the atmospheric parameter of interest, $w(n,i)$
is a weight function and $S_o(n,i)$ is the value of the processed
synthesis, each of them relative to exposure $n$ at the pixel $i$.  
$S_o$ is a function not only of the stellar parameters, but also of the 
polynomial coefficients for the flux continuum, which are determined 
uniquely for each set of stellar parameters. 
The final set of atmosphere parameters
will be the one that minimizes the $\chi^2$.
The masks and weights need to be discussed now in more detail.

\paragraph{Atmosphere parameter masks}\label{sec:atmosph-param-masks}
Although there are no lines affected only by a single parameter, every 
line reacts differently to a change of one of them. 
To determine which pixels are more sensitive to a specific parameter
compared to the others, we created several masks, applying
sequentially the following scheme for each stellar parameter:
\begin{itemize}
\item two extreme values based roughly on the expected photometric errors 
are chosen for the atmospheric parameter under analysis;
\item two syntheses with the extreme values of this parameter are
calculated, with the other parameters kept fixed to the photometric values;
\item when the difference between the two synthesis is more than 5\%
(10\% for metallicity), the binary mask is assigned a positive value,
otherwise it is declared to be null.
\end{itemize}
The ranges used for masks creation are 
$\Delta \textrm{\teff} = \pm 150\, \kelvin$,  
$\Delta \textrm{\logg}  = \pm 0.5$,   
$\textrm{\vmicro} = \pm 0.5\, \kms$, and 
$\textrm{\gfeh} = \pm 0.5$.

The metallicity mask is actually used only to exclude from the
continumm the points that are affected by a change in the overall 
metallicity (\ie\ spectral lines). 
Although the continuum is being determined as a part of every 
synthesis/observation comparison, we want to minimize as much as 
possible the influence of spectral features. 
The continuum mask has positive
value only when all the other masks have corresponding null values. 

An additional mask is created using the \ion{Fe}{1} lines from the line-list
provided by \cite{Ivans:2006hc} and \cite{Sousa:2011gr}. 
These lines have been selected for Equivalent Width measurements in high resolution 
spectra (respectively HIRES and HARPS), and their astronomical \loggf\
values have been carefully determined. 
This mask is the one used in the determination of the \gfeh\ value.

Since the phometric values of the atmospheric parameters are
reasonably close to the ones finally derived in our analyses, we
do not expect any significant change between the mask obtained with
the photometric value and the one that we would obtain if the real
atmosphere parameters were already known.

\paragraph{Weight function}
The weight function $w(n,i)$ is basically a correction factor to give
more importance to the stellar lines with respect to the sky ones in the
$\chi^2$ determination. 
For every atmospheric parameter two extreme values are selected around
the photometric value (namely, $\Delta \textrm{\teff} = \pm 200 \,\kelvin$,  
$\Delta \textrm{\logg} = \pm 0.5$,   $\Delta \textrm{\vmicro} = \pm
0.5 km/s$), 
a synthesis is generated for every combination of the parameters, and for 
every wavelength point in the synthesis the lowest and highest values from 
all the syntheses are saved. 
Hereafter  we refer to these two spectra simply as 
\textit{minimum} and \textit{maximum} spectra.

The two spectra are translated to the observational space as explained
earlier, including the addition of the sky background. 
We define a \textit{selection strip} by adding the expected Poissonian noise 
multiplied by two to the processed maximum spectrum, and subtracting the 
Poissonian noise multiplied by three to the processed minimum one. 
We use a lower factor for the upper limit because the range of
variation of a spectrum is set by its continuum, so outliers are easier to identify and a tighter
constraint can be used. 
The Poissonian noise is computed using the overall (sky and star) 
continuum level.

All the points outside the strip are excluded from the $\chi^2$ computation, 
as well as data points flagged as unusable during spectral extraction, 
by giving them null weight. 

With this procedure, we are able to safely exclude undetected lines,
anomalous pixel values (\eg\ due to anomalous measurement of flux or
inaccurate flat-field correction) and cosmic rays that survived to the
optimal extraction or have not been flagged properly. 

The weight is determined using Equation~\ref{eq:pixref}:
\begin{equation}\label{eq:pixref}
    w(i,n) = \sqrt{2  - (S_{syn}^{max}(i,n)-S_{syn}^{min}(i,n))/2}
\end{equation}
where $S_{syn}^{max}$ and $S_{syn}^{min}$ are respectively the maximum
and the minimum spectra before being processed. 
With the weight defined in this way, more importance is given to the line 
belonging only to the stellar spectrum while the global fit is preserved,
resulting in an improved atmosphere parameter determination.

\begin{figure}
\centering
\includegraphics[width=\linewidth]{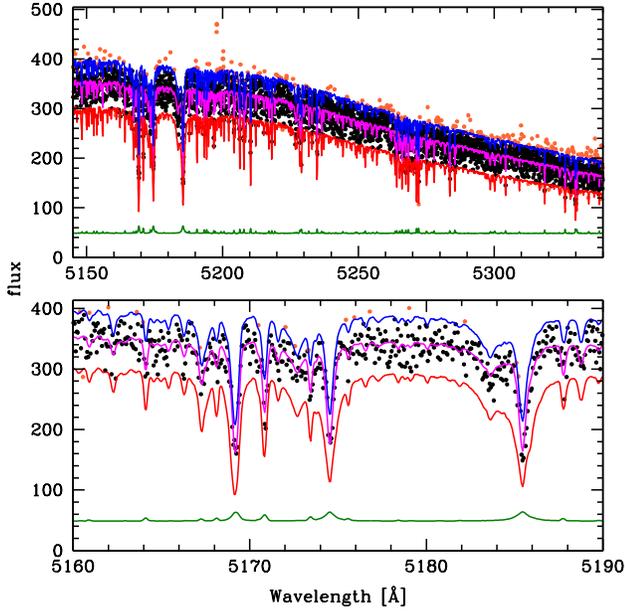}
\caption{The same spectrum in Figure~\ref{fig:spectra_faint} is
  shown (black points for the observed spectrum and magenta for the
  synthesis), with the upper boundary (blue line) and the lower one (red)
  highlighted. Points that have been excluded from the $\chi^2$
  determination are flagged in orange. 
  The green line represents the weight function $w(i,n)$,
  multiplied by a factor of 50.}
\label{fig:pixref_faint}
\end{figure}

In Figure~\ref{fig:pixref_faint} the selection strip is shown. 
Points excluded from the $\chi^2$ determination have been flagged in orange. 
A green line shows the weight function resulting from 
Equation~\ref{eq:pixref}, multiplied of a factor of 50.

We decided to perform this selection before the $\chi^2$ determination
and not during the synthesis-observation comparison, \ie\ using the
photometric values as reference instead of the actual
synthesis as a proxy, since the inclusion or exclusion of different
points or a change of the value of the weights during the $\chi^2$
determination can affect the final results.

\paragraph{$\chi^2$ minimization and atmosphere parameters determination}
Due to the complex interdependence of the atmosphere parameters, we
have opted for a simple but robust method for the $\chi^2$. 
Each iteration is characterized by a set of \textit{input} values of \teff,
\logg, \vmicro\  and \gfeh; in the first iteration the inputs are the
photometric values.
Each parameter is varied individually while the other are kept
fixed at their input values, and the value that minimize the $\chi^2$ of 
the variable parameter is saved in the \textit{output} set. 
A new set is generated by adding to each of the parameters of the input 
set half of the difference between output and input set parameters, and the 
resulting values are taken as the new input value for the next iteration. 
The cycle is broken when the difference input and output set parameters 
are $<15 \,\kelvin$ for 
temperature, $<0.05 \,\kilo\meter/\second$ for microturbulent
  velocity 
and $<0.05 ~\textrm{dex}$ for the other parameters, with all these conditions to be satisfied simultaneously. 
If convergence is not reached, determination is considered unsuccessful
and the star is dropped.

At every iteration of the algorithm a new set of atmosphere parameters is
generated, and a synthesis with these parameters must be obtained to
proceed with the $\chi^2$ minimization. 
The synthesis is calculated by linear interpolation of the pre-existing 
grid of spectra introduced in Section~\ref{sec:spectral-synthesis}. 
If the required syntheses are not included in the grid, they are generated 
at run-time and included in the grid. 
With this expedient only the synthesis needed
by the program are created, optimizing the coverage of the parameters
space against the time spent to create such grid.

\section{Results}\label{sec:results}

\subsection{Stellar atmosphere parameters and average metallicity}\label{sec:results_sub1}

The techniques described above are applied to determine the stellar
atmosphere parameters \teff, \logg, \vmicro\ and \gfeh\ of our initial
set of 2771 program stars. 
Of these stars, 2570 have reliable radial velocities, and 1910 have 
spectra of sufficient quality to determine the atmosphere parameters
without any arbitrary constraints.

\begin{figure}
\includegraphics[width=\linewidth]{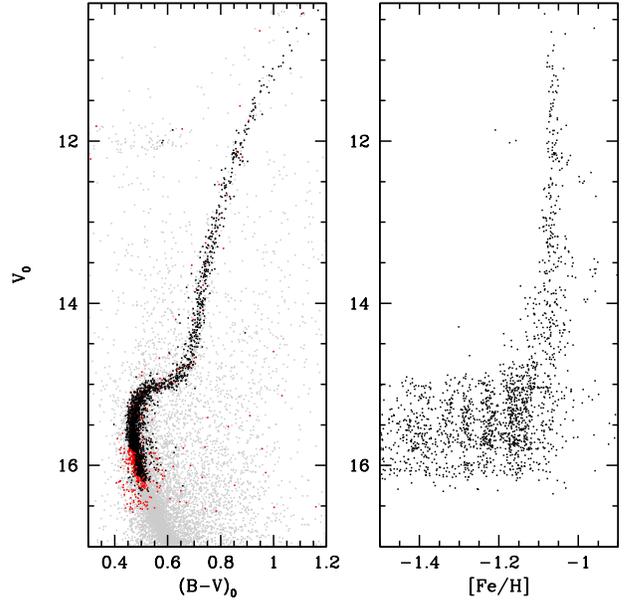}
\caption{Metallicity as a function of $V_0$ magnitude is shown
  here. 
  The average derived metallicity for RGB stars, \ie\ stars with magnitude 
  $V_0<14.7$ and color $(B-V)_0 > 0.70$, is nearly constant.
  However, fainter/bluer SGB and MS/TO stars have significantly lower
  metallicities with higher star-to-star metallicity dispersion.
  See text for discussion.
  }\label{fig:cmd_gfeh_nofixvmic}
\end{figure}

\begin{figure}
\includegraphics[width=\linewidth]{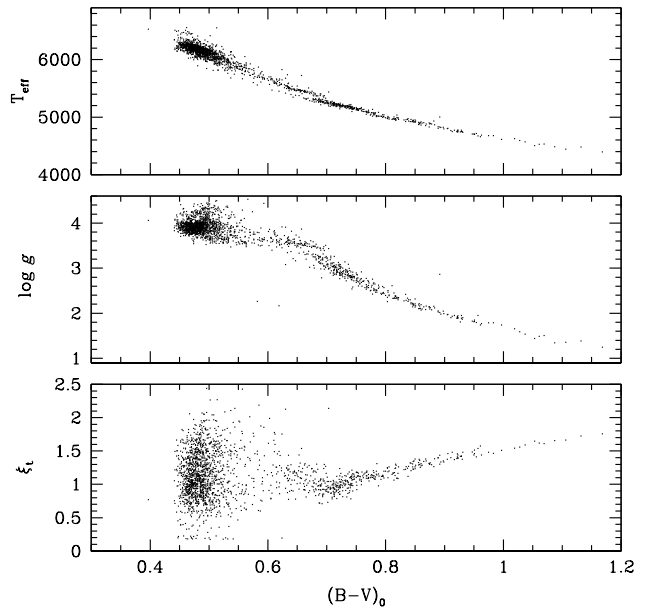}
\caption{The trend of effective temperature (upper plot), gravity
  (middle plot) and microturbulent velocity (lower plot) as a
  function of color $(B-V)_0$. The jump at  $(B-V)_0 \simeq
0.70$ is probably due to the unreliable determination of \vmicro,
which in turn is affecting the other parameters during the $\chi^2$ minimization }\label{fig:chi2mes_bvms_nofixvmic}
\end{figure}

\begin{figure}
\includegraphics[width=\linewidth]{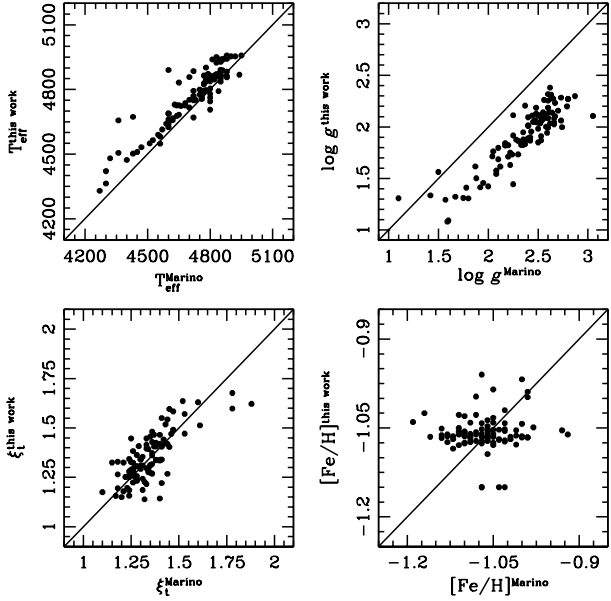}
\caption{Comparison between the atmospheric parameters and
  metallicities obtained in \citep{Marino:2008du} with the one
  obtained in this work, for the 100 stars in common between the two samples.
}\label{fig:marino_this_comp}
\end{figure}

The color-magnitude of the stars with successful determinations is
shown in Figure~\ref{fig:cmd_gfeh_nofixvmic} (left-hand panel). 
Stars for which the algorithm failed in finding the atmosphere parameters 
are marked as red. 
The trend of derived metallicity is shown in the right-hand panel. 
For stars with $V_0<14.7$ we have an average value for metallicity of
$\textrm{\gfeh} = -1.068 \pm 0.001 ~\textrm{dex}$  and a dispersion
around the mean of $\sigma = 0.025 ~\textrm{dex}$, with the quoted values referring to internal random errors only
(systematics can be as high as $0.1 ~\textrm{dex}$). 
This value, derived using 322 stars along the RGB, is in good 
agreement with $\textrm{\gfeh} = -1.07 \pm 0.01 ~\textrm{dex}$ 
determined by \cite{Marino:2008du} (M08). 
That study derived metallicities of 105 M\,4 RGB stars ($V_0 \geq 12.6$) from 
high-resolution \textrm{UVES} spectra, using EW-based abundance analyses.
A more detailed comparison of the atmospheric parameters for the 100
stars in common with M08 is shown in
Fig. \ref{fig:marino_this_comp}. The main difference (in the sense
M08 minus this study) and the relative dispersion for each
atmospheric parameter are $\Delta \textrm{\teff} = -54\ \kelvin$, $\sigma_{\Delta \textrm{\teff}}
= 62\ \kelvin$; $\Delta \textrm{\logg} = 0.45$, $\sigma_{\Delta \textrm{\logg}}
= 0.16 $;  $\Delta \textrm{\vmicro} = -0.003 \kilo\meter/\second$, $\sigma_{\Delta \textrm{\vmicro}}
= 0.088 \kilo\meter/\second$; $\Delta \textrm{\gfeh} = -0.025$, $\sigma_{\Delta \textrm{\gfeh}}
=0.090$.
Gravity is the only parameter that differs significantly from M08;
this could be explained by the lack of \ion{Fe}{2} lines in our
wavelength range. The main driver for the gravity determination in our
technique are the wings of the \ion{Mg}{1}~b triplet, which could be
affected by an incorrect value assumed for the Mg
abundance. 
Gravity however seems to have a little influence on the 
 determination of the other parameters; in fact assuming for gravity  a value derived
   with different assumed stellar masses or distances in Equation~\ref{eq:phot_gravity}
does not produce different results for the other parameters.  This is
 likely due to the high number of spectral lines belonging to neutral
 elements (mostly \ion{Fe}{1} and \ion{Ti}{1}) which are insensitive
 to gravity in our temperature range.

This overall concordance suggests that our approach to
  continuum determination and parameter estimation is reliable.

For 1588 stars fainter than $V_0=14.7$ (bluer than $(B-V)_0 \simeq 0.70$)
the derived mean metallicity decreases fairly abruptly to 
$\textrm{\gfeh}= -1.162 \pm 0.002 ~\textrm{dex}$, and the dispersion
in metallicity increases just as quickly to $\sigma=0.18 ~\textrm{dex}$. 
This behavior can be better understood by analyzing the trend of the 
other stellar parameters with color.

In Figure~\ref{fig:chi2mes_bvms_nofixvmic} we have plotted derived values
of \teff, \logg, and \vmicro\ as functions of $(B-V)_0$.
It can be seen that at bluer colors (higher \teff\ values, larger
magnitudes), \ie\ at lower \snr, the derivation of microturbulent velocity 
becomes unreliable, with a huge dispersion that is not easily explainable 
from a physical point of view. 
It is intrinsically difficult to derive \vmicro\ in M~4 MS/TO and SGB
stars because (a) the weak lines that were on the linear part of the
curve-of-growth in RGB stars now disappear into the higher
spectroscopic noise (Fig.~\ref{fig:CMD_NORED}), and (b) the lines that were strong in RGB stars
now often weaken toward the linear power curve-of-growth, losing their
sensitivity to \vmicro.
The combination of these two effects causes a reduction 
of independent data points, \ie, points that are more sensitive to one 
stellar atmospheric parameter than the others. 
The fitting procedure is not able to break the degeneracy
between \vmicro\ velocity and the other parameters, settling down to a
set of values dependent on their initial values and the path
followed by the $\chi^2$ minimization algorithm. 
This effect can be seen also in the jump in \teff\ in conjunction with the 
point where \vmicro\ stops following a defined trend at $(B-V)_0 \simeq
0.70$. 
Indeed that point corresponds to the points where the SGB joins the RGB. 

\begin{figure}
\centering
\includegraphics[width=\linewidth]{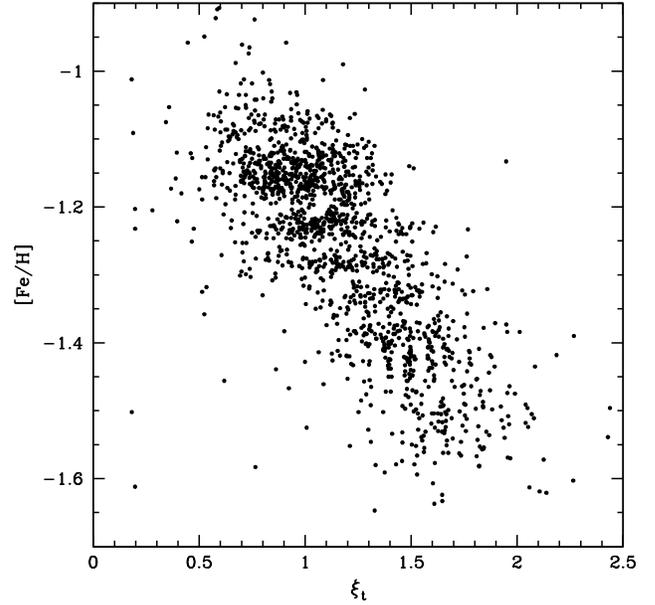}
\caption{Correlations of \gfeh\ as function of \vmicro\ 
  for stars with $(B-V)_0 < 0.7$ and $V_<14.7$.
  A strong trend of metallicity with microturbulent velocity
  is present for these MS/TO and SGB  stars.}
\label{fig:oth_vs_gfeh_nofixvmic}
\end{figure}

Incorrect derivations of stellar parameters of course affects
the determination of metallicity, and Figure~\ref{fig:oth_vs_gfeh_nofixvmic}
shows the clear correlation of \gfeh\ with the microturbulence
velocity for MS/TO and SGB stars. 
Unfortunately there appears to be no reliable calibration 
in the literature of \vmicro\ as a function of colors and magnitude for
these kinds of stars.

\begin{figure}
\includegraphics[width=\linewidth]{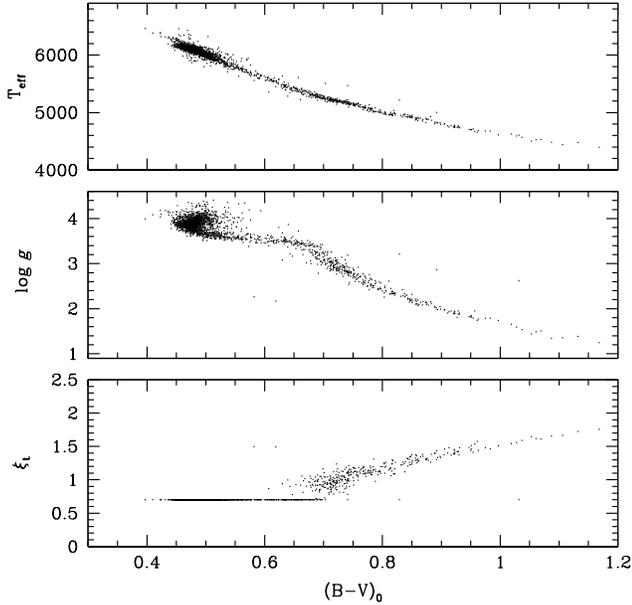}
\caption{Atmospheric parameters after combining giants with \vmicro\
  as a free parameter and dwarfs with fixed \vmicro. The transition
  between the two groups in temperature and gravity is now smoother.}\label{fig:chi2mes_bvms_fixvmic}
\end{figure}

We decided to fix at $\textrm{\vmicro}=0.7 \,\kilo\meter/ \second$ the
microturbulence for stars with  $V_0 > 14.7 $. 
This value has been derived according to the trend of \vmicro\ defined by 
the RGB stars, extrapolated to the SGB and MS/TO regime.
With this constraint, the number of stars successfully analyzed grows from 1910 
stars to 2191, of which there are 322 RGB and 1869 SGB and MS/TO stars.
The resulting atmosphere parameters are shown in 
Figure~\ref{fig:chi2mes_bvms_fixvmic}. 
The transition between the two groups is now smooth.

\begin{figure}
\includegraphics[width=\linewidth]{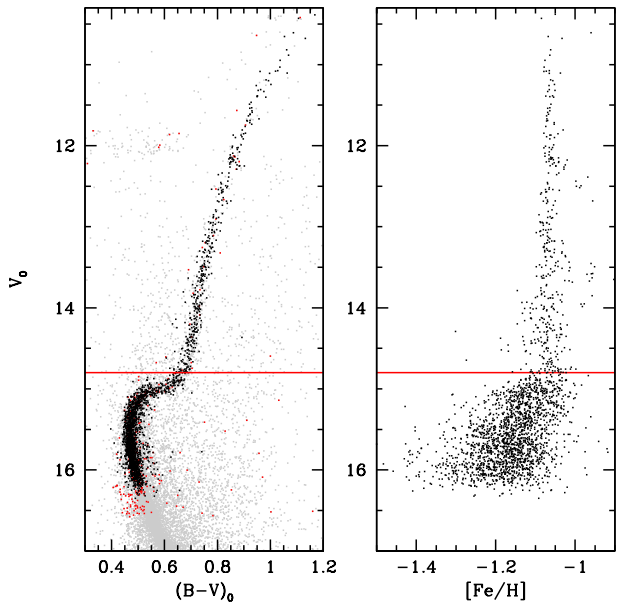}
\caption{Metallicity as a function of $V_0$ magnitude is shown
  here. For MS/TO and SGB stars a microturbulent velocity of
  $\textrm{\vmicro}=0.7 \,\kilo\meter/\second$ has been assumed (stars
  under the red line). 
  The average metallicity for MS and SGB stars,  
  $\textrm{\gfeh}= -1.16 ~\textrm{dex}$, is significantly
  lower than   RGB stars. 
Assuming different values for \vmicro\ does not
  reduce significantly this difference. }\label{fig:cmd_gfeh_fixvmic}
\end{figure}

\begin{figure}
\includegraphics[width=\linewidth]{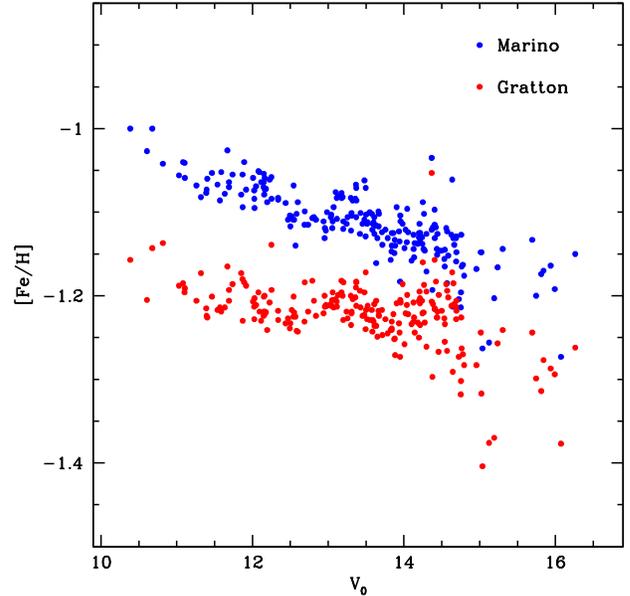}
\caption{Resulting metallicities when using Equations
  ~\ref{eq:micro_marino} (blue points) and ~\ref{eq:micro_gratton}
  (red points). The presence of trends and systematics offsets with
  literature values convinced us to leave the microturbulent velocity
  as a free parameter for giant stars}\label{fig:vmic_vs_gfeh}
\end{figure}

Before opting to use a fixed value for microturbulent
  velocity, we have checked if Equations
  ~\ref{eq:micro_marino} or ~\ref{eq:micro_gratton} could be used to
  fix this parameter for all the stars in the sample. Figure\ref{fig:vmic_vs_gfeh}
  shows that when using Equation~\ref{eq:micro_marino} (blue points)
  the derived metallicities have a
  marked trend with magnitude already at the level of RGB stars, while
  the metallicities derived by using Equation~\ref{eq:micro_gratton}
  (red points) are systematically lower than the ones obtained either
  from equivalent width analysis or when leaving the microturbulent
  velocity as a free parameter for giant stars.  

Finally the trend of metallicity with magnitude is shown in
Figure~\ref{fig:cmd_gfeh_fixvmic}. The mean metallicity measured from MS and
SGB stars is significantly lower than the one measured from RGB
stars. From 1869 stars with $V_0>14.7$, we have an average metallicity
of $\textrm{\gfeh}= -1.162 \pm 0.002 ~\textrm{dex}$, with a dispersion of
$\sigma=0.09 ~\textrm{dex}$.  
We have tested several values for the microturbulence velocity and we
find that a positive variation of $0.1$ \textrm{dex} in \vmicro\ causes a net
change of $\simeq -0.01$ \textrm{dex} in the derived metallicity. Thus
it cannot explain the $0.10$ difference between the S/SGB and the RGB samples.

Whether this metallicity trend is a consequence of 
the difficulties we have in extracting abundances at these very low SNR 
levels, or a real result (as suggested for other clusters,
see, e.g., \citealt{Lind:2008dg}) shall be further investigated.

Our overall metallcity scale for M4 is dependent on a 
large number of choices that we have made in reduction of observed spectra, 
generation of synthetic spectra, and the matching of them.
In earlier sections of this paper we have discussed our analytical 
decisions in detail.
However, as emphasized in, \eg, \cite{Kraft:2003a}, all
abundance analyses are tied to their adopted methodologies.
In Table~\ref{tab-m4metal} we gather many of the M\,4 metallicity
estimates from the literature.
This list is probably not complete, and concentrates on more recent
\gfeh\ values.
Neglecting the early small-sample study of \cite{Geisler:1984a}, 
the mean of the remaining 18 literature values is $<$\gfeh$>=-1.18$,
with $\sigma = 0.08$; the extreme values range from $-1.05$ to $-1.32$.
Our derived values of $-$1.07 (SGB/RGB stars) and $-1.16$ (MS/SGB
stars) are consistent with the literature.
Note that \cite{Monaco:2012li} derive metallicities for their M4
MS stars that are 0.14~dex smaller than those of their SGB/RGB sample.
This is in the same sense and about the same magnitude as we find 
between our MS/SGB and SGB/RGB stars.

Although we still suspect that analytical difficulties attend the
derivation of reliable metallicites for our low \snr\ spectra of the
MS/SGB group, it is possible that other factors may be contributing
to the small mean \gfeh\ shift seen here. More investigations on
this point are required.

\subsection{$\alpha$-elements and metallicity}
To test the influence of  $\alpha$-elements
  on the
  determination of metallicity, we have varied all the abundances listed
  in section~\ref{sec:spectral-synthesis} by $\pm 0.1~\textrm{dex}$
  and performed again our analysis  on a subset of stars randomly
  selected along the sequences in the CMD.

\begin{figure}
\includegraphics[width=\linewidth]{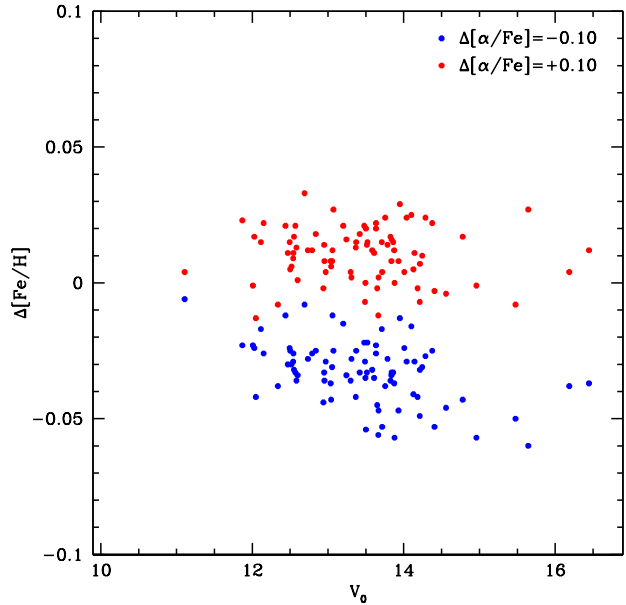}
\caption{The influence of the assumed abundances for
    $\alpha$-elements (section \ref{sec:spectral-synthesis}) on the
    determination of metallicity is tested by varying the quoted values of $\pm 0.1 ~\textrm{dex}$ and determining again the atmospheric parameters and metallicity. }
\label{gfeh_vs_alpha}
\end{figure}

Figure~\ref{gfeh_vs_alpha} shows that a $ +0.1
  \textrm{dex}$ variation in alpha elements corresponds to an average
  increase in \gfeh of $ 0.01 ~\textrm{dex}$, while the same variation
  of $\alpha$-elements in the opposite direction is causing a decrease in metallicity of 
$\pm 0.03~\textrm{dex}$.
Since the difference between the true abundances of
  $\alpha$-elements and the ones derived by \cite{Ivans:1999hf} is
  likely smaller
  than the variation we have used to test their influence on
  metallicity, and being this difference one order of magnitude
  smaller than other systematics effects, we can safely state that our
  results are independent of the exact choice of $\alpha$-elements abundances.

\subsection{Metallicity of the multiple sequences}\label{sec:results_sub1}

The existence of two sequences in the RGB of M\,4 has been
demonstrated spectroscopically by M08 and subsequently observed photometrically by
\cite{Monelli:2013us} (MO13) on a larger sample of stars by using an
appropriate combination of photometric filters. To determine if there
is any difference in metallicity between the two sequences, we have
cross-matched their selection (upper panels of Figure 4 in MO13) with
our catalogue. Here we are retaining the same color code and naming
convention used in MO13 to distinguish the two sequences, \ie green
for RGBa and magenta for RGBb. The cross-match results in 99 stars for
RGBa and 207 stars for RGBb. 

\begin{figure}
\includegraphics[width=\linewidth]{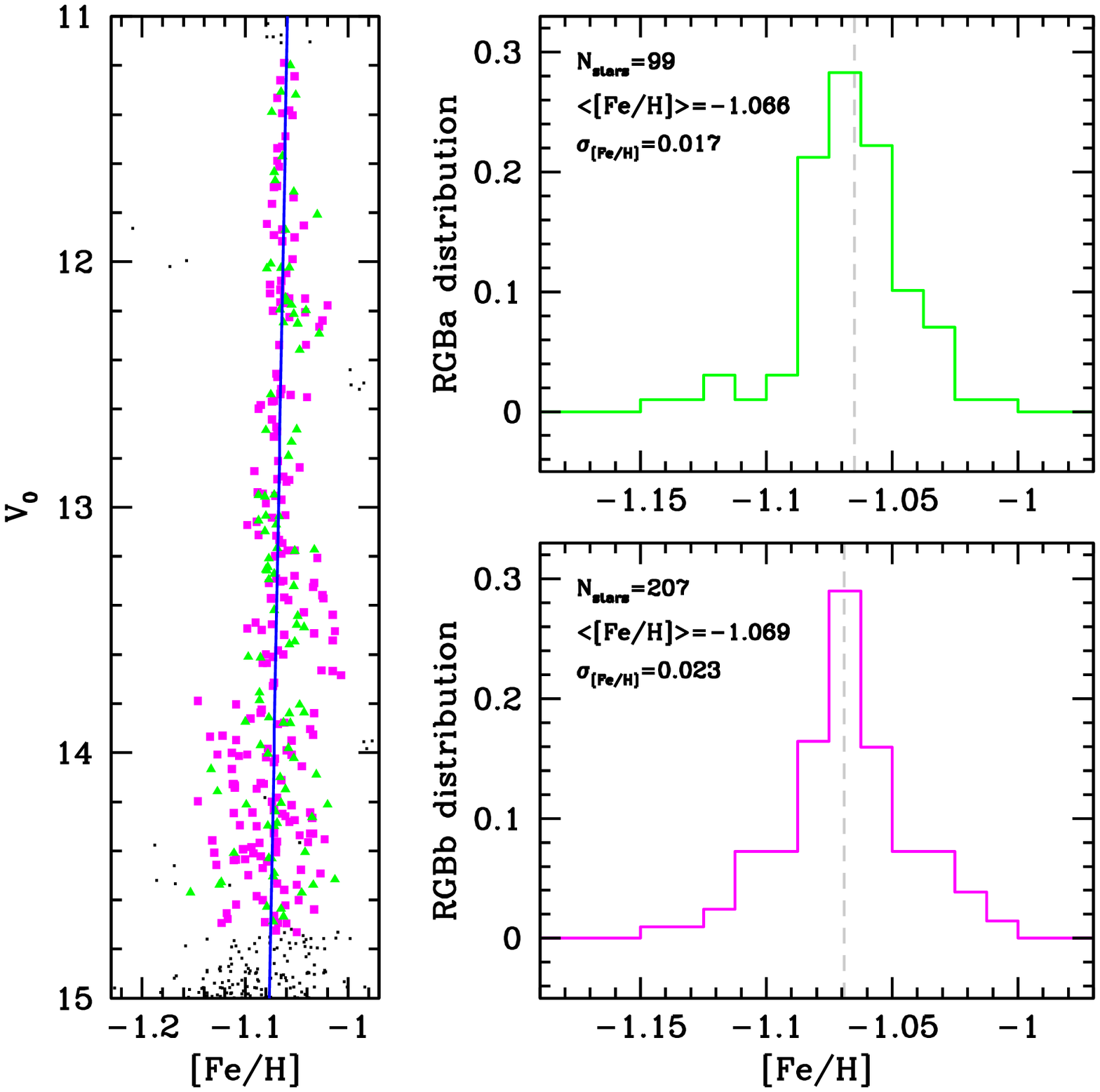}
\caption{
{\it Left} Color-magnitude diagram resulting from the cross-match of our sample with
  \cite{Monelli:2013us}  database for the two RGB sequences. 
We used
the same color code and naming
convention (green for RGBa and magenta for RGBb) 
as in \cite{Monelli:2013us}. 
The blu line
represents the slope use to empirically remove the trend of
metallicity with magnitude. 
{\it Right.} Normalized distribution in \gfeh for the two stellar populations.
The gray lines represent
the median of each distribution.}
\label{fig:gfeh_monelli}
\end{figure}

Before determining the mean metallicity of each sequences, we have decide to 
empirically correct for linear trends of \gfeh\ with magnitude to
better highlights systematic differences between the two
sequences. By fitting all the stars in the RGB (blue line in the
left panel of Figure~\ref{fig:gfeh_monelli}) we have found a linear
trend of $-0.0044~\textrm{dex}/\textrm{mag}$ with the zero point at
$V_0=13.0,~\textrm{\gfeh}=-1.0677$. Both RGBa and RGBb stars have been corrected
for the same trend before the determination of the mean metallicity. 
From our data it is not possible to determine if this trend is caused by a systematic effect of model atmosphere with temperature or by our analysis technique, \eg a non-perfect continuum normalization. This trend however is so tiny that it does not influence our previous discussion.
  
The histograms of the two normalized distributions are shown in the left panels
of Figure~\ref{fig:gfeh_monelli}. We find a median metallicity of
$\textrm{\gfeh}^{RGBa} = -1.066$ with a dispersion (determined as the
68\textsuperscript{st} of the residuals from the median) of
$\sigma^{RGBa}_{\textrm{\gfeh}} = 0.017$ for the sequence RGBa, and
$\textrm{\gfeh}^{RGBb} = -1.069$ with $\sigma^{RGBb}_{\textrm{\gfeh}}
= 0.023$ for the second sequence. With a  Spearman's rank correlation
coefficient of $-0.19$ and an error on the mean metallicity of each
sequence lower than $0.002~\textrm{dex}$, we
conclude that no difference in metallicity between the two sequences is visibile at a
level of $0.01~\textrm{dex}$.

\section{Conclusions}\label{sec:conclusions}

We have presented here a new algorithm to automatically determine the continuum normalization
level for spectra characterized by a very short and almost
continuum-less wavelength range, very low \snr\ and strong
contamination from sky background. We have coupled this algorithm with a
modified $\chi^2$ minimization algorithm for the derivation of stellar
parameters by comparison with spectral syntheses. Data points have
been weighted according to the sensitivity of spectral features with
respect to variations in the atmosphere parameters. 

The robustness of our algorithm has 
been tested over 322 RGB stars down to $V\simeq 16$ resulting in 
an almost
constant metallicity along the entire RGB and an average value of
$\textrm{\gfeh} = -1.068 \pm 0.001 ~\textrm{dex}$, consistent with
the one derived in \cite{Marino:2008du} from 105 high-resolution
\textrm{UVES} spectra of bright stars ($V<14$) and using Equivalent
Width analysis. We stress the fact that not only the same results are
recovered despite using observations  with lower spectral resolution and a
wavelength range that is almost  
ten times shorter, but we are able to extend the analysis to stars that are two
magnitudes fainter than the ones in the high-resolution sample. 
We have not  found any significant difference in metallicity at the
level of $0.01 ~\textrm{dex}$ between the two 
RGB sequences identified by M08 and MO13.

At lower magnitudes atmosphere parameter determination has proven to
be more difficult due to 
the faintness of the stars. We fixed the microturbulent velocity at
$\textrm{\vmicro}=0.7 \,\kilo\meter/\second$ to remove one degree of
freedom in the $\chi^2$ minimization.  From 1869 stars in the Sub
Giant Branch and Main Sequence we have an average metallicity 
of $\textrm{\gfeh}= -1.162 \pm 0.002 ~\textrm{dex}$, with a dispersion
of $\sigma=0.09 ~\textrm{dex}$. Assuming slightly different values of
microturbulent velocity does not remove the observed difference in
metallicity. A decrease in surface metallicity around the
Turn-Off has been already observed in other globular clusters 
and attributed to diffusive processes (sedimentation, levitation)
coupled with  turbulent mixing below the outer convection, see ad
example \citealt{Lind:2008dg}. Due to the tricky characteristics of the spectra in
this temperature range and the low \snr\ involved, we have decided to perform more
tests before confirming this result.

M4 is known to have a nearly constant [Fe/H] metallicity along 
the RGB \citep{Ivans:1999hf,Marino:2008du,Carretta:2009kr}.
Thus it has proved to be the best candidate to identify systematic effects
introduced by our technique in its initial development phase. 
Now that many critical issues have been identified and solved, we are
planning to apply our method to other GCs that show interesting
features in their upper RGB or that have multiple sequences in their CMD. 
NGC\,6752 has been targeted for an RV survey similar to the one of
M4 \citep{Milone:2006cc} so it represents the next natural target.
Other GCs like M\,22, NGC\,6397, 47\,Tuc and
$\omega$ Centauri have a wealth of data in the GIRAFFE archive; they
present interesting targets for analysis in the future.

\acknowledgments
We thank Dr. Peter Stetson for providing the photometric catalogue 
and Melike Af{\c s}ar for helpful discussions.
Partial support for this work has been provided by the US National Science
Foundation under grants AST-0908978 and AST-1211585. LM, GP, and VN
recognize partial support by the Universita' degli Studi di Padova CPDA101477 grant.
LM acknowledges the financial support provided by \textit{Fondazione
Ing. Aldo Gini}. APM acknowledges the financial support from the Australian Research
Council through Discovery Project grant DP120100475. We thank the anonymous
referee for the thorough review and for providing useful suggestions
that improved the manuscript.

\begin{center}
\begin{deluxetable}{rrr}
\tablewidth{0pt}
\tablecaption{Spectroscopic Observation Statistics\label{tab-nobs}}
\tablecolumns{3}
\tablehead{
\colhead{No. Observations}    & 
\colhead{No. Stars}           & 
\colhead{Total Spectra}       
}
\startdata
  1   &       83     &      83     \\
  2   &     1722     &    3444     \\
  3   &      501     &    1503     \\
  4   &      321     &    1284     \\
  5   &       72     &     360     \\
  6   &       20     &     120     \\
  7   &        1     &       7     \\
  8   &       27     &     216     \\
  9   &       10     &      90     \\
 10   &       12     &     120     \\
 11   &        1     &      11     \\
 12   &        1     &      12     \\
      &              &             \\
Total &     2771     &    7250     \\
\enddata
\end{deluxetable}
\end{center}

\begin{center}
\begin{deluxetable}{cc}
\tablewidth{0pt}
\tablecaption{Reddening Parameters for M\,4\tablenotemark{a}\label{tab-parameters}}
\tablecolumns{2}
\tablehead{
\colhead{Quantity}    &  
\colhead{Value}       
}
\startdata
Distance     &   $1.80 \pm 0.05 \kilo \textrm{pc}$      \\
$A_V$        &   $ 1.39 \pm 0.01$                       \\
$ (M-m)_0$   &   $11.18 \pm 0.06$                       \\
$E(B-V)$     &   $0.37 \pm  0.01$                       \\
$E(V-I)$     &   $0.53 \pm 0.01 $                       \\
$R_V$        &   $ 3.76 \pm  0.07$                      
\enddata
\tablenotetext{a}{\cite{Hendrick:2012cw}}
\end{deluxetable}
\end{center}

\begin{center}
\begin{deluxetable}{lccc}
\tablewidth{0pt}
\tablecaption{Metallicity Results for M\,4\label{tab-m4metal}}
\tablecolumns{4}
\tablehead{
\colhead{Reference}    &  
\colhead{No. Stars}    & 
\colhead{\gfeh}        & 
\colhead{Sample}       
}
\startdata
  \cite{Geisler:1984a}          & $-0.94$   &       2    &  RGB                  \\
  \cite{Zinn:1984cc}            & $-1.28$   &  $-$       &  metallicity scale    \\
  \cite{Gratton:1986a}          & $-1.32$   &       3    &  RGB                  \\
  \cite{Zinn:1984cc}            & $-1.20$   &       4    &  RGB                  \\
  \cite{Lambert:1992ie}         & $-1.11$   &       2    &  BHB                  \\
  \cite{Brown:1992a}            & $-1.21$   &       3    &  RGB                  \\
  \cite{Drake:1994a}            & $-1.05$   &       4    &  RGB                  \\
  \cite{Carretta:1997ht}        & $-1.19$   &       3    &  RGB                  \\
  \cite{Ivans:1999hf}           & $-1.18$   &      36    &  RGB                  \\
  \cite{Kraft:2003a}            & $-1.18$   &  $-$       &  metallicity scale    \\
  \cite{Yong:2008b}             & $-1.24$   &      14    &  RGB                  \\
  \cite{Marino:2008du}          & $-1.07$   &     105    &  RGB                  \\
  \cite{Carretta:2009kr}        & $-1.20$   &     103    &  RGB                  \\
  \cite{Carretta:2009id}        & $-1.17$   &      14    &  RGB                  \\
  \cite{Mucciarelli:2011jr}     & $-1.10$   &      87    &  MS to RGB            \\
   \cite{Villanova:2012ls}      & $-1.14$   &      23    &  RGB                  \\
  \cite{Monaco:2012li}          & $-1.31$   &      71    &  MS                   \\
  \cite{Monaco:2012li}          & $-1.17$   &      10    &  SGB/RGB              \\
  \cite{Villanova:2012a}        & $-1.06$   &       6    &  BHB                  \\
  this study                    & $-1.07$   &     332    &  SGB/RGB              \\
  this study                    & $-1.16$   &    1869    &  MS/SGB               \\
\enddata
\end{deluxetable}
\end{center}

\end{document}